\documentclass[%
 reprint,
 amsmath,amssymb,
 aps,
 pra,
]{revtex4-1}

\newcommand{\comment}[1]{}
\usepackage[english]{babel}
\usepackage{float}
\usepackage{siunitx}
\usepackage{xcolor}

\usepackage{graphicx}% Include figure files
\usepackage{dcolumn}% Align table columns on decimal point
\usepackage{bm}% bold math
\begin{document}

%\preprint{APS/123-QED}

\title{Spectroscopy and ion thermometry of C$_{2}^{-}$ using laser-cooling transitions}

\author{Markus N{\"o}tzold}
\author{Robert Wild}
\author{Christine Lochmann}
\author{Roland Wester}%
\email{roland.wester@uibk.ac.at}
\affiliation{%
 Institut f{\"u}r Ionenphysik und Angewandte Physik, Universit{\"a}t Innsbruck, Technikerstra{\ss}e 25, 6020 Innsbruck, Austria
}%

\date{\today}% It is always \today, today,

\begin{abstract}
A prerequisite for laser cooling a molecular anion, which has not been achieved so far, is the precise knowledge of the relevant transition frequencies in the cooling scheme. To determine these frequencies we present a versatile method that uses one pump and one photodetachment light beam. We apply this approach to C$_{2}^{-}$ and study the laser cooling transitions between the electronic ground state and the second electronic excited state in their respective vibrational ground levels, $B ^{2} \Sigma _{u} ^{+}(v=0) \leftarrow X ^{2} \Sigma _{g} ^{+}(v=0) $. Measurements of the R(0), R(2), and P(2) transitions are presented, which determine the transition frequencies with a wavemeter-based accuracy of $0.7\times10^{-3}$\,cm$^{-1}$ or 20\,MHz. The spin-rotation splitting is resolved, which allows for a more precise determination of the splitting constants to $\gamma '  = 7.15(19)\times10^{-3}\,$cm$^{-1}$ and $\gamma '' = 4.10(27)\times10^{-3}$\,cm$^{-1}$. These results are used to characterize the ions in the cryogenic 16-pole wire trap employed in this experiment. The translational and rotational temperature of the ions cooled by helium buffer gas are derived from the Doppler widths and the amplitude ratios of the measured transitions. The results support the common observation that the translational temperature is higher than the buffer gas temperature due to collisional heating under micromotion, in particular at low temperatures. Additionally, a rotational temperature significantly lower than the translational is measured, which agrees with the notion that the mass weighted collision temperature of the C$_{2}^{-}$-He system defines the internal rotational state population.
\end{abstract}

                              
\maketitle

\section{Introduction}

Nowadays laser cooling is a routine technique utilized in many laboratories to gain precise control over neutral and positively charged species. However, laser cooling has not been achieved for negatively charged atoms or molecules. This is due to the loosely bound excess electron in an anion which makes the system more fragile in comparison to cations. Hence, the basic requirement of multiple electronic states for efficient laser cooling are rarely met, one exception being dipole-bound states~\cite{Lykke1984:prl,Carelli2014:jcp,Yuan2020:prl,Simpson2021:prl}. Since the energies of dipole-bound states are very close or even above the binding energy of the excess electron, a dissociation of the anion via autodetachment becomes likely. Consequently, these states are not feasible for a cooling scheme.

There are a few atomic anions with more than one stable electronic states below the detachment threshold~\cite{Warring2009:prl,Jordan2015:prl,Tang2019:prl}, and most notably the molecular anion C$_{2}^{-}$. Due to the high electron affinity of the neutral C$_{2}$ molecule of about 3.3\,eV~\cite{Jones1980:jcp,Ervin1991:jpc}, the anion is able to suppport three stable electronic states. In addition, the branching ratios of Franck-Condon factors are favourable~\cite{Shan-Shan2003:cp,Shi2016:ctc}, making C$_{2}^{-}$ a promising candidate for laser cooling a negative ion~\cite{Yzombard2015:prl}. A laser cooled anion would enable sympathetic cooling of other anionic species, including antiprotons~\cite{Gerber2018:njp}, which is also a proposed method to produce cold anti-hydrogen~\cite{Baker2021:n}.

The molecule C$_{2}^{-}$ is among the most researched anions theoretically~\cite{Shi2016:ctc,Barsuhn1974:jpb,Zeitz1979:cpl,Dupuis1980:jcp,Rosmus1984:jcp,Nichols1987:jcp,Watts1992:jcp,Sedivova2006:mp,Kas2019:pra,Gulania2019:fd}. Also experimentally it has a long spectroscopic history with first transitions observed by Herzberg and Lagerqvist in 1968~\cite{Herzberg1968:cjp}. In the following decades
$B ^{2} \Sigma _{u} ^{+} \leftarrow X ^{2} \Sigma _{g} ^{+}$ 
and 
$A ^{2} \Pi _{u} ^{+} \leftarrow X ^{2} \Sigma _{g} ^{+}$ 
transitions were studied by several groups with increasing precision and for different vibrational levels using fluorescence spectroscopy, photodetachment spectroscopy and photoelectron spectroscopy~\cite{Jones1980:jcp,Mead1985:jcp,Ervin1991:jpc,Rehfuss1988:jcp,Lineberger1972:cpl,Royen1992:jms,Beer1995:jpc,Bragg2003:cpl,Shan-Shan2003:cp, Nakajima2017:jms}.
Furthermore, the molecular anion C$_{2}^{-}$ has also been the focus of storage ring experiments, studying radiative lifetimes~\cite{Pedersen1998:jcp} and more recently radiative cooling mechanisms~\cite{Iida2020:jpcl}. Atomic carbon anions have been studied by photodetachment following core-hole excitation~\cite{Perry-Sassmannshausen2020:prl}. C$_{2}^{-}$ is iso-electronic to N$_2^+$, which holds promise for precise vibrational spectroscopy experiments \cite{Germann2014:np}.
It has also been suggested that C$_{2}^{-}$ should be present in the interstellar medium due to the abundance of neutral C$_{2}$~\cite{Lambert1995:aj,Souza1977:aj,Lambert1986:ajss}, but evidence of its presence is still awaiting confirmation~\cite{Civis2005:pasj}.

Although there are many studies on C$_{2}^{-}$, the reported term values lack the necessary precision for laser cooling. In this work we achieve a crucial step towards this goal by a precise determination of laser cooling transitions in the  $B ^{2} \Sigma _{u} ^{+} \leftarrow X ^{2} \Sigma _{g} ^{+}$ system with an accuracy of about 20\,MHz. This is slightly larger than the natural linewidth of about 2\,MHz, corresponding to a lifetime of the excited state of 75\,ns~\cite{Leutwyler1982:cpl,Rosmus1984:jcp}. The studied transitions are used to derive not only the translational, but also the rotational temperature of the ion ensemble in our newly developed cryogenic wire trap.

\section{Methods}

%%%%%%%%%%
\begin{figure*}[t]
\centering
\includegraphics{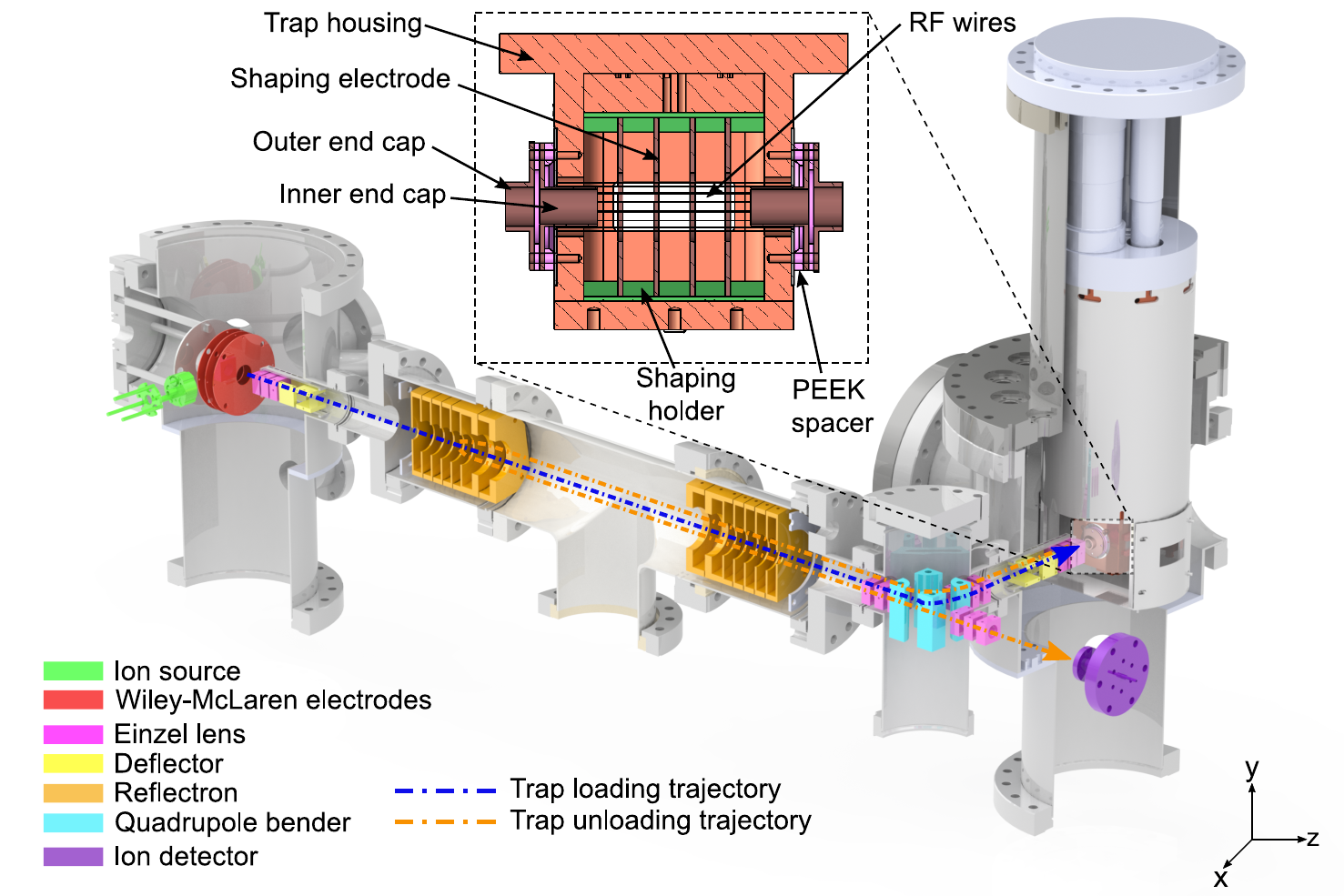}
\caption{Schematic of the experimental setup. The most important components, in particular the ion optics elements, are highlighted in distinct colors. The two dashed lines depict the trajectory of the ion ensemble when loading the ions into the trap~(blue) or after extraction from the trap~(orange). The inset shows a section view of the ion trap.}
\label{fig:setup}
\end{figure*}
%%%%%%%%%%

%%%%%%%%%%
\begin{figure*}[t]
\centering
\includegraphics{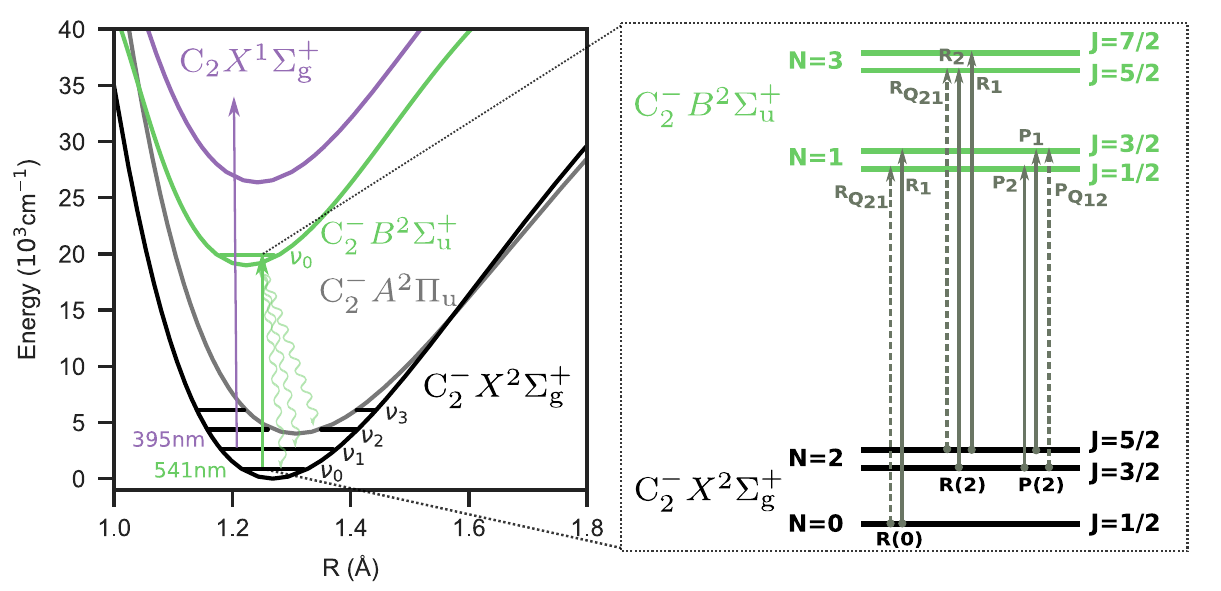}
\caption{The left panel shows the potential energy curves of the C$_{2}^{-}$ system. The right panel depicts the 541nm transition $B ^{2} \Sigma _{u} ^{+}(v=0) \leftarrow X ^{2} \Sigma _{g} ^{+}(v=0) $ on the scope of rotational levels including spin-rotation splitting. Due to the zero spin nuclei of C$_{2}^{-}$, the electronic ground state only has even rotational levels $N$, while the second excited state only has odd rotational levels.}
\label{fig:potential_energy_curves}
\end{figure*}
%%%%%%%%%%

The spectroscopic measurements rely on determining a loss of ions stored in the trap when irradiated by a light beam that matches the resonance of an electronic transition. The trap in which the C$_{2}^{-}$ anions are stored is a linear radio-frequency~(rf) multipole wire trap mounted on a cryostat kept at a temperature of roughly 6\,K. The temperature can be adjusted by adding a heat load to the cryostat, which in our case is achieved via heater resistors located in the trap housing. The trap design is similar to the wire trap described in Ref.~\cite{Geistlinger2021:rsi}.
A schematic of the experimental setup, including a section view of the ion trap, is shown in Figure~\ref{fig:setup}. The rf trap used in the experiment consists of 16 wires, with diameters of 100\,\si{\micro\meter}, that are arranged on a circle of 1\,cm diameter.
The sinusoidal rf drive voltage applied for trapping the C$_{2}^{-}$ ions has an amplitude of 150\,V and a frequency of 7.3\,MHz.
Axial confinement is created via two hollow end caps that have an inner diameter of 7\,mm, are 0.5\,mm thick, and are a distance of 31.5\,mm apart.
The static voltage applied to the inner end caps is -3\,V. The shaping electrodes have an inner diameter of 14\,mm and are 1\,mm thick. They can be used to further modify the shape of the ion cloud, however, for the present measurements they are kept at ground potential.

For the measurement scheme two distinct light beams are used. The photodetachment beam has a power of roughly 330\,mW and is created 
%by collimating light 
from a UV-LED~(IN-C39ATOU5) operating at 395\,nm. The laser beam driving the electronic transitions is created via a DL pro laser from Toptica at 1082\,nm which is frequency doubled to 541\,nm through a crystal~(MSHG 1080) from Covesion. To record and lock the laser frequency, the wavelength meter Angstrom WS Ultimate 2 by HighFinesse is used. During the measurements, the wavelength meter is calibrated several times using a diode laser that is locked on a rubidium D line with Doppler-free saturation spectroscopy. From this we estimate a systematic error on the wavelength measurements of 10\,MHz or $0.34\times10^{-3}$\,cm$^{-1}$. For all the presented measurements the pump beam is passed into the trap through the hollow end caps, i.\ e.\ in axial direction, with a continuous power of up to 10\,\si{\micro\watt}. The photodetachment beam is coupled into the trap from a side window. 

The C$_{2}^{-}$ ions are created from a plasma discharge source using acetylene seeded in argon and are selectively loaded into the trap via time-of-flight~\cite{Best2011:aj}. Once loaded into the trap, the anions' rotational and translational temperature thermalize with a helium buffer gas at a density of about 10$^{14}$\,cm$^{-3}$. The ions are then exposed to a 541\,nm laser beam for 10\,s, driving the transition $B ^{2} \Sigma _{u} ^{+}(v=0) \leftarrow X ^{2} \Sigma _{g} ^{+}(v=0) $ as depicted in the left panel of Figure~\ref{fig:potential_energy_curves}. The various rotational transitions including the line splitting due to spin-rotation coupling are depicted in the right panel of Figure~\ref{fig:potential_energy_curves}. From the $B ^{2} \Sigma _{u} ^{+}(v=0)$ state the anions spontaneously decay to different vibrational levels of the electronic ground state. As a final step the 395\,nm beam is added into the trap, which neutralizes anions in $v=1$ and higher via the process of photodetachment $\text{C}_{2}^{-} + h\nu \xrightarrow{} \text{C}_{2} + \text{e}^{-}$. Afterwards the remaining ions are extracted axially and are destructively measured with a microchannel plate detector. Due to the large number of ions, about a few hundred to few thousand, the detector does not resolve individual ions but produces an electron current proportional to the amount of ions that strike the detector. This is fully sufficient for the present experiments as we only need to analyze relative changes in ion signal.

To derive a quantity proportional to the transition rate from the measured ion signal we need to calculate the relative population of the ions in the vibrational ground state after exposure to the 541\,nm beam~($t=0$\,s). The measured ion signal as a function of the photodetachment beam exposure time is described as 
\begin{equation}
a(t) = C_{0}\; \text{e}^{-k_{0} t} + C_{1} \; \text{e}^{-k_{1} t}.
\end{equation}
The variables $ C_{0}$ and $C_{1}$ are the amplitudes of the anions in the vibrational ground and excited state in $X ^{2} \Sigma _{g} ^{+}$. The lifetimes $1/k_{\text{0}}=659(19)$\,s and $1/k_{\text{1}}=24.4(3)$\,s are due to the exposure of the photodetachment beam~(395\,nm) and are determined in separate measurements. Note that the vibrational ground state C$_{0}$ has a finite loss rate due to the spectral width of the UV-LED. To obtain $k_{\text{0}}$, a measurement omitting the 541\,nm pump beam was performed, such that all stored ions remain in the vibrational ground state. For determining $k_{\text{1}}$, the ions are exposed to the pump beam~(541\,nm) permanently, which results in an ion ensemble in $v=1$ and higher. The optical elements controlling beam diameters and overlap with the ions are optimised to maximize the loss rate $k_{\text{1}}$. To obtain a relative population of the vibrational ground state $C_{\text{g}} = \frac{C_{0}}{C_{0}+C_{1}}$ we measure the ion signal at $t=0$\,s and $t=70$\,s, and solving for $C_{\text{g}} $ yields
\begin{equation}
C_{\text{g}} = \frac{\text{e}^{k_{0} 70\mathrm{s}}   (a(0\mathrm{s})-a(70\mathrm{s}) \; \text{e}^{k_{1} 70\mathrm{s}})}
{a(0\mathrm{s}) \; (\text{e}^{k_{0}  70\mathrm{s}} - \text{e}^{k_{1} 70\mathrm{s}})} .
\end{equation}
As the $B ^{2} \Sigma _{u} ^{+}$ state is short lived with a 75\,ns lifetime~\cite{Leutwyler1982:cpl,Rosmus1984:jcp}, it can be safely assumed that no electronically excited anions are present due to the comparatively long experimental time scales. Since the lifetime of the trapped ions embedded in the helium buffer gas is about 20 hours, it can also be excluded from the analysis. The relative population of the ground state is then connected to the wavelength-dependent transition rate $k(\lambda)$ via
\begin{equation}
\label{eq:pump_rate}
C_{\text{g}} = \text{e}^{-k(\lambda)P\rho 10\mathrm{s}},
\end{equation}
where $P$ is the incoming photon flux and $\rho$ is the geometric overlap between the 541\,nm laser beam and the ion cloud. Thus, we can calculate a quantity proportional to the transition rate via $ k(\lambda)  \propto -\ln ( C_{\text{g}} )/P$,
which in following figures is plotted as relative signal. As the overlap between the three-dimensional ion density distribution and the laser beam cannot be exactly determined, only the laser power is used for normalization. The geometric overlap is assumed to be constant. The error bars on the relative signal in the figures represent the 1-sigma uncertainties derived from the statistical fluctuations of the recorded ion signal, laser power, and time constants $k_0$ and $k_1$. Fluctuations in the laser-ion overlap would also appear as fluctuations of the signal and are thus incorporated in the statistical error.

So far it was assumed that our initial ion ensemble has a thermal distribution of the rotational levels and that the populations of the rotational $N$ and vibrational $v$ levels are then modified due to the pump laser~(541\,nm). However, in the experiment, the $N$ levels are constantly redistributed due to collisions with the helium buffer gas. Note that this effect is negligible for the vibrational levels as the cross section for state-changing collisions between C$_{2}^{-}$ and helium has been calculated to be many orders of magnitude smaller for vibrations than for rotations.~\cite{Mant2020:ijms,Mant2020:pra}. If the optical transition rate is higher than or comparable to the redistribution rate, then the measurement will show a saturation effect when scanning over the transition. To circumvent this problem, the power of the 541\,nm beam is reduced to a level of below a microwatt, such that the rate of the ions optically pumped into higher vibrational levels is small compared to the rate at which the rotational levels are redistributed.

\begin{figure}[b]
\centering
\includegraphics{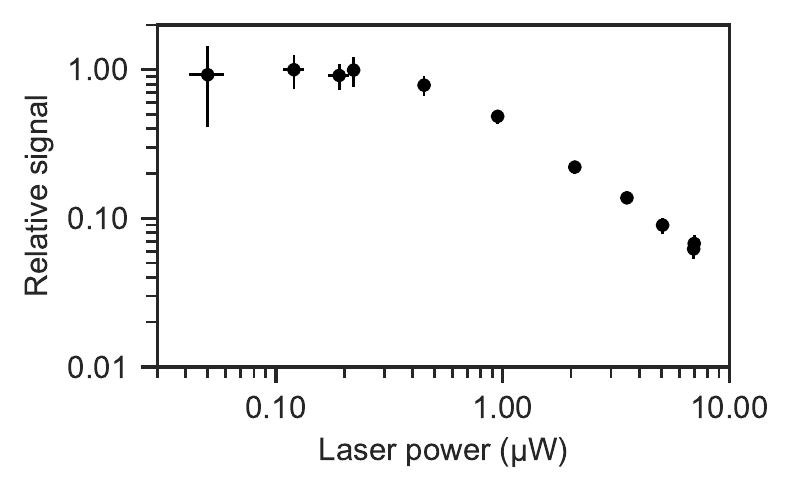}
\caption{Relative signal calculated via Equation~(\ref{eq:pump_rate}) for the center of the R(0) transition as a function of the 541\,nm laser beam power.}
\label{fig:pump_power_dependance}
\end{figure}

The finite timescale for rotational redistribution causes an underestimation of the measured transition rate at too high laser powers as can be seen for the main peak of the R(0) transition in Figure~\ref{fig:pump_power_dependance}. Only for laser powers below about 0.5\,$\mu$W a constant plateau emerges, meaning that in this region the rotational redistribution is faster than the laser excitation rate. The timescale for rotational redistribution can be estimated from the buffer gas density and the calculated rotationally inelastic collision rate coefficients \cite{Mant2020:ijms} to about 100\,$\mu$s. Thus, for about 0.5\,$\mu$W a similar timescale for optical excitation can be expected. For the measurements presented, the laser power is adjusted so that we operate below this power in the region of the plateau. To obtain similar signal-to-noise ratios, the powers are varied between 0.3 and 0.4\,\si{\micro\watt}.

%%%%%%%%%%
%%%%%%%%%%

\section{Results and Discussion}
\subsection{Spectroscopy of C$_{2}^{-}$}

%%%%%%%%%%
\begin{figure}[b]
\centering
\includegraphics{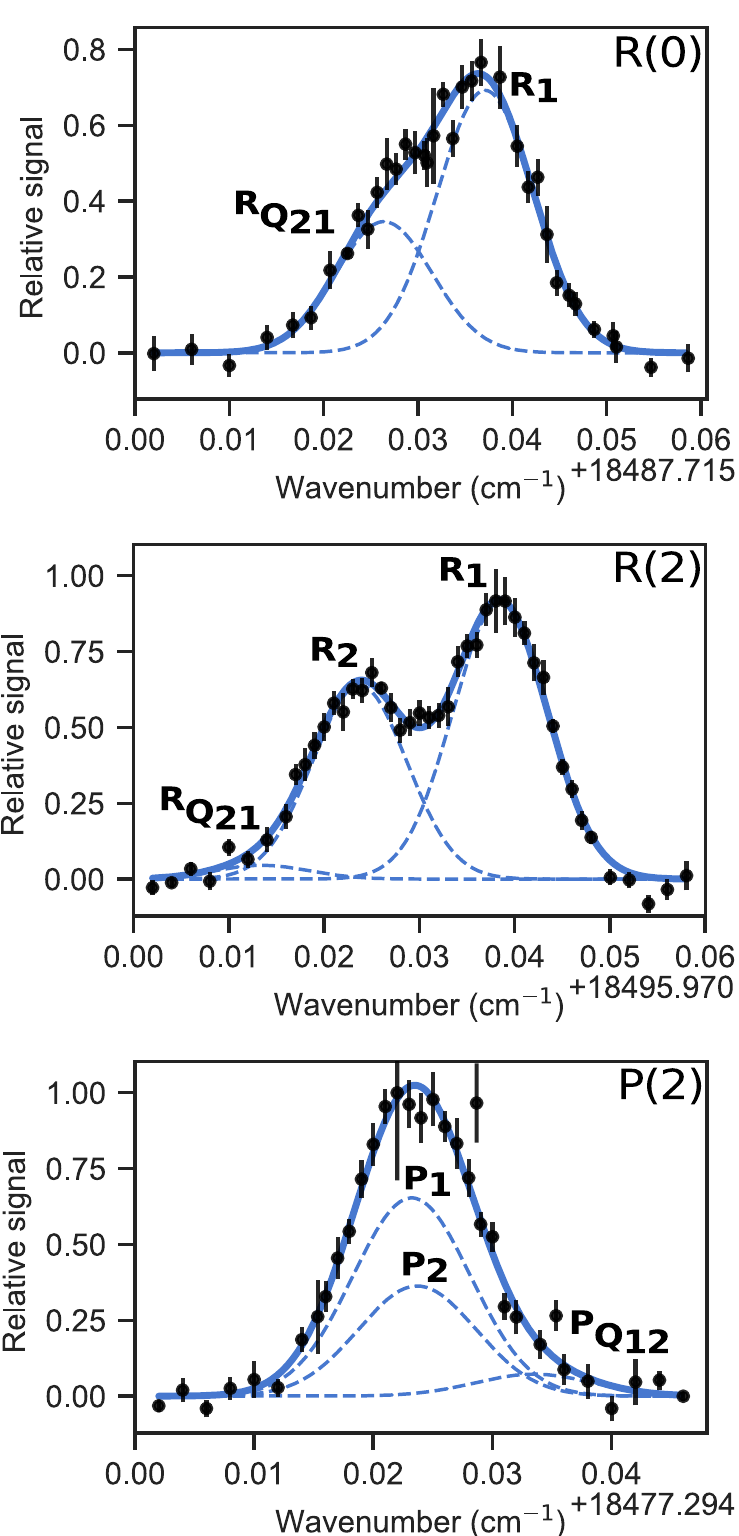}
\caption{Spectra of R(0), R(2) and P(2) which are transitions of the $B ^{2} \Sigma _{u} ^{+}(v=0) \leftarrow X ^{2} \Sigma _{g} ^{+}(v=0) $ band. The dashed lines show individual Gaussians described by Equation~(\ref{eq:doppler_gauss_fit}), the solid line is the fit function defined as the sum of dashed functions. The different plots use the axis offset given below the graphs, respectively.}
\label{fig:transitions}
\end{figure}
%%%%%%%%%%

Applying the measurement scheme described in the previous section, we were able to measure the three rotational transitions R(0), R(2) and P(2) of $B ^{2} \Sigma _{u} ^{+}(v=0) \leftarrow X ^{2} \Sigma _{g} ^{+}(v=0) $ which can be seen in Figure~\ref{fig:transitions}. In the figure two distinct peaks in the R(2) transition can be seen and a shoulder in R(0). This is caused by the line splitting of the rotational transitions which occurs because the electron spin couples to the rotation of the molecule in two ways, either increasing or decreasing the total angular momentum. Consequently, two different energy levels exist per rotational level $N$, except for $N=0$. The additional energy levels allow for two distinct transitions from $N=0$ and three distinct transitions from higher $N$ values, also illustrated in the right panel of Figure~\ref{fig:potential_energy_curves}. The naming convention of the transitions is taken from Ref.\ \cite{Herzberg1950}.

To describe the fine structure of the absorption profiles we fit them with the sum of either two (for R(0)) or three (for R(2) and P(2)) Gaussians. The fit function using the optimized fitting parameters is shown by the solid line in Figure~\ref{fig:transitions}, while the contribution of individual Gaussians is depicted via dashed lines. Each Gaussian is defined as
\begin{equation}
\label{eq:doppler_gauss_fit}
    f(\nu)=S_{J} A\; \text{e}^{ -\frac{m c^{2}} {k_{\text{B}}T} ( \frac{\nu - \nu_{\text{c}}}{\nu_{\text{c}}})^{2} }.
\end{equation}
Here, $S_{J}$ is the H{\"o}nl-London factor describing the intensity of the rotational transition, $A$ is a scaling parameter for the amplitude, $m$ is the mass of C$_{2}^{-}$, $c$ the speed of light, $k_{\text{B}}$ the Boltzmann constant, $T$ the temperature, and $\nu_{\text{c}}$ the center frequency of the transition. The Gaussians have individual $\nu_{\text{c}}$ parameters, but share $A$ and $T$ for each transition. The scaling parameter $A$ can be shared because the relative amplitudes of the individual functions are fixed via the theoretical line strength $S_{J}$. The value of $S_{J}$ depends on the initial and final rotational level and is described by one the following equations~\cite{Herzberg1950}
\begin{subequations}
\begin{align}
  S_{J}^{R}(J'')&=\frac{ (J''+1)^{2} - \frac{1}{4} } {J''+1},
 \\
  S_{J}^{Q}(J'')&=\frac{ 2J''+1 } {4J''(J''+1)},
 \\
  S_{J}^{P}(J'')&=\frac{ J''^{2} - \frac{1}{4} } {J''}.
\end{align}
\end{subequations}

The observed transition frequencies of $B ^{2} \Sigma _{u} ^{+}(v=0) \leftarrow X ^{2} \Sigma _{g} ^{+}(v=0) $ for R(0), R(2) and P(2) are summarized in Table~\ref{tbl:transitions} along with previous measurements~\cite{Royen1992:jms}. Besides the statistical error given in the table, the systematic error of $0.34\times10^{-3}$\,cm$^{-1}$ from the wavelength meter calibration needs to be included. Based on the sum of the two errors, the combined overall accuracy of the measured transitions is better than $0.7\times10^{-3}$\,cm$^{-1}$ or 20\,MHz. Comparing the measured transition frequencies with work from~\cite{Royen1992:jms} shows a systematic offset of roughly 0.01cm$^{-1}$, which is attributed to the finite accuracy of the previous measurements.

\begin{table}[b]
\caption{\label{tbl:transitions}
Measured rotational transitions of $B ^{2} \Sigma _{u} ^{+}(v=0) \leftarrow X ^{2} \Sigma _{g} ^{+}(v=0) $ given in cm$^{-1}$. The statistical error and the systematic error from the wavelength meter calibration are given separately. Note that Ref.~\cite{Royen1992:jms} does not give error estimates of the measured values.
}
\begin{ruledtabular}
\begin{tabular}{ccc}
Transition & This Work\,(cm$^{-1}$) &
Ref.~\cite{Royen1992:jms}\,(cm$^{-1}$) \\
\hline

        $R_{1}(0)$ & 18487.75238(19)(34) & 18487.764\\
        
        $^{R}Q_{21}(0)$ & 18487.74165(28)(34) & \\
        
        $R_{1}(2)$ & 18496.00878(11)(34) & 18496.020\\
        
        $R_{2}(2)$ & 18495.99402(17)(34) & 18496.010\\
        
        $P_1(2)$,$P_2(2)$ & 18477.31728(14)(34) & \\

\end{tabular}
\end{ruledtabular}
\end{table}    

From the partially resolved R(0) and R(2) transitions the spin-rotation splitting constants $\gamma'$ and $\gamma''$ can be derived, which correspond to the electronically excited state $B ^{2} \Sigma _{u} ^{+}(v=0)$ and ground state $X ^{2} \Sigma _{g} ^{+}(v=0)$, respectively. The splitting constant $\gamma'$ is directly obtained from the R(0) spectrum. It is then fed as a fixed parameter into the R(2) fit function, to fix the relative distance of the $^{R}Q_\text{21}$ transition. Then, $\gamma''$ is determined from the distance between the two visible peaks in R(2).

More explicitly, we first define the rotational term values~\cite{Herzberg1950}
\begin{subequations}
\label{eq:rotationanl_term_values}
\begin{align}
    F_{1}(K) &= B_{v} N(N+1) + \frac{1}{2} \gamma N,
    \\
    F_{2}(K) &= B_{v} N(N+1) - \frac{1}{2} \gamma (N+1).
\end{align}
\end{subequations}
Here $B_{v}$ is the rotational constant of vibrational level $v$, $\gamma$ the spin-rotation splitting constant and $N$ the rotational level quantum number. The subscript 1 defines levels of $J=N+\frac{1}{2}$, subscript 2 levels of $J=N-\frac{1}{2}$. Using these term values the four main branches are described by
\begin{subequations}
\begin{align}
    R_{1}(N) &= \nu_{0} + F_{1}'(N+1) - F_{1}''(N),
\\
    R_{2}(N) &= \nu_{0} + F_{2}'(N+1) - F_{2}''(N),   
\\
    P_{1}(N) &= \nu_{0} + F_{1}'(N-1) - F_{1}''(N),
\\
    P_{2}(N) &= \nu_{0} + F_{2}'(N-1) - F_{2}''(N).   
\end{align}
\end{subequations}
Here $v_0$ is the frequency of the transition without taking rotations into account. Further, the the two satellite branches are described via
\begin{subequations}
\label{eq:last_branch}
\begin{align}
    ^{R}Q_{21}(N) &= \nu_{0} + F_{2}'(N+1) - F_{1}''(N),    
\\
    ^{P}Q_{12}(N) &= \nu_{0} + F_{1}'(N-1) - F_{2}''(N).    
\end{align}
\end{subequations}

The spectrum of R(0) is fitted with a superposition of two Gaussians with the free fitting parameters being $A$, $T$, $\nu_\text{c}$, and $d_0$, where $d_0$ defines the distance between the two Gaussians.
Using equations~(\ref{eq:rotationanl_term_values})-(\ref{eq:last_branch}) we then calculate the splitting constant $\gamma'$ from $d_{0} =^{R}Q_{21}(0) - R_{1}(0) = -\frac{3}{2}\gamma'$.
For the R(2) spectrum, the fit function is a superposition of three Gaussians with the free fitting parameters being $A$, $T$, $\nu_\text{c}$, $d_1$, and the constraint $d_2$. Here, $d_1$ defines the distance between two main peaks seen in the spectrum and  $d_2$ uses $\gamma'$ as a parameter to fix the distance of the satellite transition relative to $R_{1}(2)$ via $d_{2}=^{R}Q_{21}(2)-R_{1}(2)=-\frac{7}{2}\gamma'$.

From the fit of R(2) we use the distance between the two peaks of the main branch and the previously obtained value of $\gamma'$ to calculate $\gamma''$ via $d_1 =R_{1}(2)-R_{2}(2)=-\frac{7}{2}\gamma'-\frac{5}{2}\gamma''$. For the P(2) transition no individual peaks can be resolved, but it is still fitted with a superposition of three Gaussians with the free fitting parameters $A$, $T$, $\nu_\text{c}$, and the two constraints $d_3$ and $d_4$. 
Using $\gamma'$ and $\gamma''$ obtained from the other two transitions we fix the distances between the various branches as $d_3 = P_{1}(2)-P_{2}(2)=\frac{3}{2}\gamma'-\frac{5}{2}\gamma''$ and $d_4 = ^{P}Q_{12}(2)-P_{1}(2)=\frac{5}{2}\gamma''$, which yields a consistent fit of the P(2) transition.

The calculated splitting-constants $\gamma'$ and $\gamma''$ corresponding to the  $B ^{2} \Sigma _{u} ^{+}(v=0)$ and $X ^{2} \Sigma _{g} ^{+}(v=0) $ state are shown in Table~\ref{tbl:spin_splitting_constants}. Both values are in good agreement with previous measurements~\cite{Royen1992:jms, Rehfuss1988:jcp}, but with an improved precision by about a factor of two.

\begin{table}[b]
\caption{\label{tbl:spin_splitting_constants}
Spin-rotation splitting constants $\gamma'$ and $\gamma''$ in units of cm$^{-1}$, corresponding to the electronic $B ^{2} \Sigma _{u} ^{+}(v=0)$ and $X ^{2} \Sigma _{g} ^{+}(v=0) $ state of C$_{2}^{-}$.}
\begin{ruledtabular}
\begin{tabular}{ccc}
 Splitting constant & This work\,(cm$^{-1}$) & Previous meas.\,(cm$^{-1}$)\\
\hline

$\gamma' \times 10^{-3} $ & 7.15(19)  & 7.26(56)~\cite{Royen1992:jms} \\ %7.26(56) *10^3 cm^-1
$\gamma'' \times 10^{-3} $ & 4.10(27)  & 4.25(56)~\cite{Rehfuss1988:jcp} \\ %4.25(56) *10^3 cm^-1

\end{tabular}
\end{ruledtabular}
\end{table}

\subsection{Ion trap thermometry}

\begin{table}[h]
\caption{\label{tbl:temperatures}
Measured translational and rotational temperature for trapped $C_{2}^{-}$ ions, buffer gas cooled by helium. The cryostat temperature defining the helium temperature is 6.3(3)\,K.}
\begin{ruledtabular}
\begin{tabular}{cc}

    Transition & Translational temperature (K) \\
    \hline
        
    $R(0)$ & 19.8(16)   \\ 
    $R(2)$ & 18.7(7)    \\ 
    $P(2)$ & 18.5(11)   \\ 
        
    \toprule
    Amplitude ratio & Rotational temperature (K) \\
    \hline
        
    ${R(0)}/{R(2)}$ & 6.6(6)   \\ 
    ${R(0)}/{P(2)}$ & 6.7(8)   \\

\end{tabular}
\end{ruledtabular}
\end{table}

From the measured transitions the rotational as well as the translational temperature can be derived for trapped $C_{2}^{-}$ ions, which are buffer-gas cooled by helium atoms at a cryostat temperature of 6.3(3)\,K. The results are summarized in Table~\ref{tbl:temperatures}. The translational temperature of the ion ensemble is measured via the Doppler width of the various transitions and is directly obtained as a fitting parameter. For the fits of the Doppler width no line broadenings are included as the only mechanism deemed relevant is the laser line width of 2\,MHz and the natural linewidth of the transition of 2\,MHz~\cite{Leutwyler1982:cpl,Rosmus1984:jcp}, both of which are negligible compared to an expected Doppler full width at half maximum of about 200\,MHz at 6\,K.

The rotational temperature can be determined from the fits' scaling parameters $A$. The ratio of the scaling parameters of two transitions starting from different $N$ levels equals the ratio of the populations of the corresponding $N$ levels. The population distribution of the $N$ levels for a temperature $T$ is given by the partition function via
\begin{equation}
   p(N) = (2N+1)\cdot \text{e}^{-B_0^{''} N(N+1) /k_{\text{B}} T }.
\end{equation}
Hence, taking the ratio of the partition function for rotational levels $N=0$ and $N=2$ and solving for temperature $T$ we obtain
\begin{equation}
    T = \frac{6 B_0^{''} h} {k_{\text{B}} \ln \left(5 \frac{p(0)}{p(2)} \right)},
\end{equation}
where the ratio $p(0)/p(2)$ is the observable ratio of the scaling parameters $A$ in the experiment and $B_0^{''}=1.74$\,cm$^{-1}$ the rotational constant calculated via $R_{1}(0)-P_{1}(2)=6B_0^{''}$. Note that for simplicity the spin-rotation splitting and rotational distortion are ignored, since their effects are negligible for the calculation of the rotational temperature.

\begin{figure*}
\centering
\includegraphics{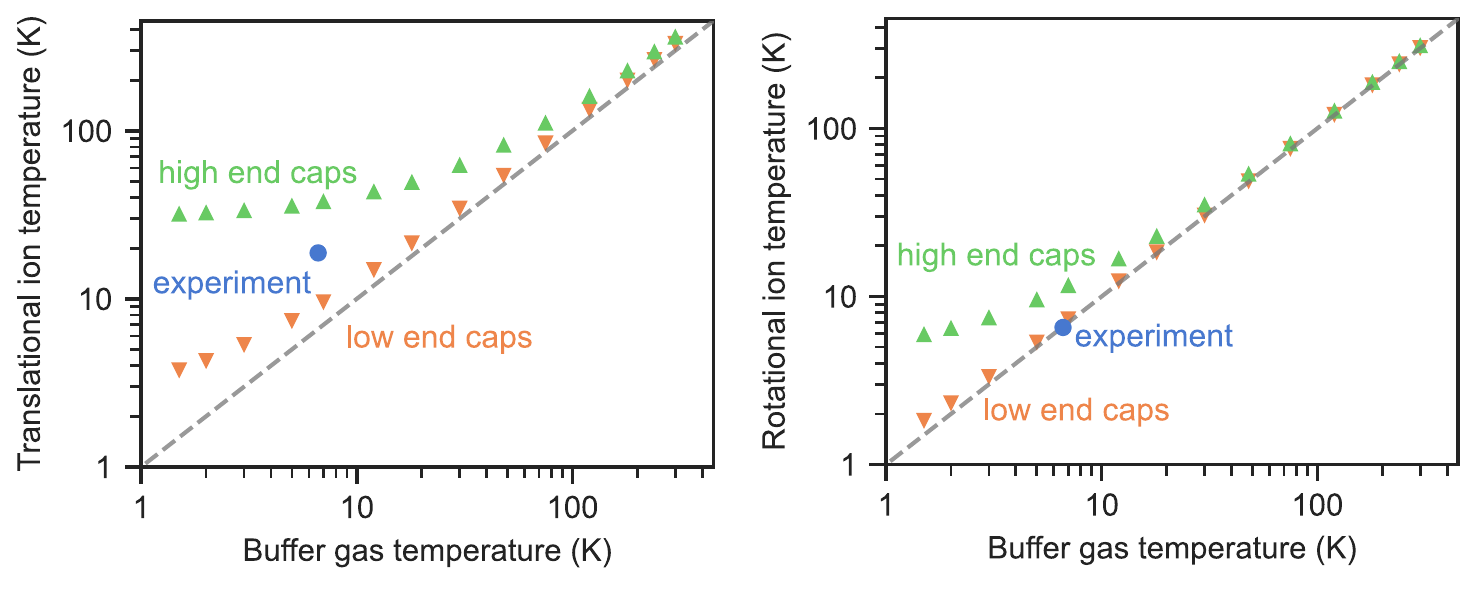}
\caption{Left panel shows the translational, right panel the rotational temperature of C$_{2}^{-}$ anions colliding with helium buffer gas. The simulated rotational temperature is calculated according to Equation~(\ref{eq:collision_temperature}). The colors in the plots correspond to an end cap voltage of -3\,V~(orange) or -30\,V~(green) of the wire trap. The experimental points~(blue) were taken with an end cap voltage of roughly -3\,V.}
\label{fig:simulations}
\end{figure*}

From the measurements we find a translational ion temperature of about 20\,K, significantly larger than the buffer gas temperature of roughly 6\,K. This increased temperature is caused by heating collisions of the ions with the buffer gas in the rf field, which results in an increased kinetic energy in the radial direction. This is repeatedly observed in rf traps theoretically~\cite{Asvany2009:ijms,Noetzold2020:as,Hoeltkemeier2016:prl} and experimentally~\cite{Jusko2014:prl,Domenech2018a:aj,Domenech2018b:aj,Lakhmanskaya2020:pra}. We can reproduce this heating effect in a molecular dynamics simulation using a 3D model of the wire trap~(see Appendix~\ref{app:md} for details). This is shown in the left panel of Figure~\ref{fig:simulations}. Note that we use the notion of a temperature even though the resulting energy distributions in the simulations show a deviation from the Maxwell-Boltzmann distribution in the form of a high energy tail~\cite{Rouse2017:prl,Noetzold2020:as}.

%%%%%%%%%%
%%%%%%%%%%

An interesting feature is that the translational temperature increasingly breaks away from the buffer gas temperature at lower energies, which is correlated to the existence of local potential minima in the radio-frequency fields~\cite{Otto2009:jpb}. These minima arise since the end caps create an axially confining potential in the middle of the trap pushing the ions radially outwards. In a quadrupole ion trap this effect weakens the radial harmonic potential. In a multipole trap without a harmonic potential this causes a radial shift outwards, and in an ideal multipole trap creates a minimum in the shape of a ring. However, due to symmetry breaking, a discrete number of local minima is observed in practice~\cite{Otto2009:jpb,Pedregosa-Gutierrez2018:rsi,Marchenay2021:qst}. The depth of these minima varies with different parameters such as the voltages applied to the end caps as well as geometrical imperfections of the trap and patch potentials, caused e.\ g.\ by impurities frozen to the electrodes. Imperfections and impurities are neglected in the simulations, as are Coulomb forces between the particles of the ion cloud, which may also result in an additional heating caused by the ions being pushed deeper into the rf field.

%%%%%%%%%%
\begin{figure}[b]
\centering
\includegraphics{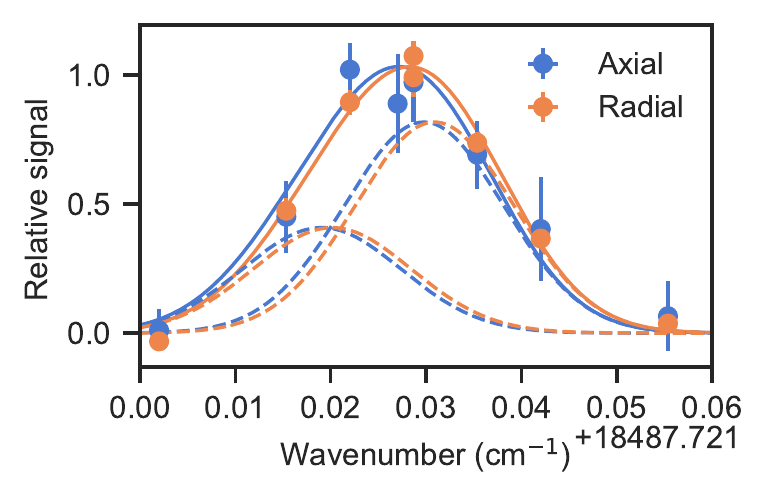}
\caption{ Measurements of the R(0) transition at a buffer gas temperature of 30\,K. The two measurements are taken with the pump laser coupled into the trap either axially~(blue) or radially~(orange). The dashed lines are individual Gaussians described by Equation~(\ref{eq:doppler_gauss_fit}), the solid line is the sum of the corresponding dashed lines. The plot uses the axis offset given below the graph.}
\label{fig:axial_radial_30K}
\end{figure}
%%%%%%%%%%

From the simulations, which assume an ideal wire trap, we find that the heating of the ions is only found in the radial direction as the rf field is negligible in the axial direction. However, in the experiment the excess energy is efficiently coupled into the axial motion. This can be seen by Doppler width measurements of the R(0) transition performed at a cryostat temperature of 30\,K (see Figure~\ref{fig:axial_radial_30K}). Here, two measurements are shown, one with the pumping light aligned axially~(through the hollow end caps), and one with the light aligned radially~(from a side window through gaps in the wires). The same fitting routine as described previously is used, but the spin-splitting constant is not a free parameter and uses the value obtained from the higher resolution measurement. Both measurements yield the same temperature of about 50\,K, also showing again a significant deviation from the buffer gas temperature. We assume that the efficient coupling between axial and radial motion is due to geometrical imperfections and stray electric fields.

For the internal degrees of freedom we derived a significantly lower temperature of about 6.6\,K compared to 19\,K of the translational temperature. This is in line with the notion that the internal rotational temperature is described by the mass weighted collision temperature
\begin{equation}
\label{eq:collision_temperature}
    T = (m_{i}T_{b} + m_{b}T_{i}) / (m_{i}+m_{b}).
\end{equation}
Here $m_{i}$ and $T_{i}$ are the mass and translational temperature of the ion C$_{2}^{-}$, while the variables $m_{b}$ and $T_{b}$ correspond to the same quantities for the helium buffer gas. It is assumed that the neutral buffer gas thermalizes with the cryostat walls which were measured to hold a temperature of 6.3(3)\,K. Our simulations of the rotational ion temperature shown in the right panel of Figure~\ref{fig:simulations} show how this temperature closely follows the buffer gas temperature. A significant deviation only occurs for very large discrepancies between the translational ion and buffer gas temperature below 3\,K.

The fact that we reach low collision temperatures in our wire trap has already been demonstrated in previous work, where we investigated temperature sensitive three-body reactions~\cite{Wild2021:jpca} down to 10\,K. In measurements using a 22-pole trap, we found similar translational temperatures of OH$^{-}$ anions~\cite{Lakhmanskaya2020:pra}, but the OH$^{-}$ rotational temperatures were not found to follow equation (11), but remained similarily high as the translational temperatures~\cite{Otto2013:pccp,Endres2017:jms}. A possible explanation might be the high trap offset voltage used for the 22-pole trap combined with grounded plates located nearby. Simulations suggest that this creates a single local minimum in the rf field, in which Coulomb repulsion could lead to additional translational heating. Another possibility might be that the permanent dipole moment of OH$^{-}$ couples more strongly than anticipated to black-body radiation that reaches inside the cryogenic ion trap.

\section{Conclusion}

We measured the R(0), R(2), and P(2) transitions of the $B ^{2} \Sigma _{u} ^{+}(v=0) \leftarrow X ^{2} \Sigma _{g} ^{+}(v=0) $ band, which has a natural linewidth of 2\,MHz, with an accuracy of about 20\,MHz. This is suitable for potential future laser-cooling applications. We also resolved the spin-rotation line splitting in both the R(0) and R(2) transitions and determined the splitting constants with improved precision over previous work.

In the experiment, helium buffer gas at 6.3(3)\,K was used to thermalize the trapped ions. This resulted in a translational temperature of 18.8(7)\,K, derived from the Doppler width of the transitions. The internal rotational temperature was determined to be 6.6(5)\,K via the ratio of the transition strengths R(0)/R(2) and R(0)/P(2). This result supports the notion that the rotational temperature is defined by the collision temperature, i.\ e.\ the mass weighted temperature of the C$_{2}^{-}$-He system.

The presented measurement scheme only requires a single diode laser for driving the respective transitions as well as a photodetachment beam created by a UV-LED for neutralizing anions. It can be further extended to transitions of higher vibrational levels by using different photodetachment beams, e.\ g.\ a 405\,nm beam created by a Blu-ray diode for detaching $v=2$ and higher. 
The measurement scheme presented here is able to measure transitions required for laser-cooling C$_{2}^{-}$ with sufficient precision, which we demonstrated by measuring the cooling transitions between the electronic ground state $X ^{2} \Sigma _{g} ^{+}$ and the excited state $B ^{2} \Sigma _{u} ^{+}$ of C$_{2}^{-}$.

As a next step, this method can be used to measure the transition frequencies of repump laser beams. By exciting ions from vibrationally excited levels in the ground electronic state into the $B ^{2}\Sigma _{u} ^{+}(v=0)$ state they may spontaneously decay back to the vibrational ground level of the $X ^{2} \Sigma _{g} ^{+}$ state. Thus, one would observe a decrease of the relative ion loss induced by UV photodetachment. The method is also applicable for measuring transitions to the $A ^{2} \Pi _{u} ^{+}$ state of C$_2^-$.

\begin{acknowledgments}
We thank Barry Mant and Francesco A. Gianturco for many helpful discussions. This work was supported in parts by the Austrian Science Fund (FWF) through Project I3159-N36.

%The column width is: \the\columnwidth
%The column width is: 7.50673589809 cm , use 7.5
%Double column width is: 15.0134717962 cm , use 15
\end{acknowledgments}

\appendix
\section{Details on the molecular dynamics simulations}
\label{app:md}
%%%%%%%%%%
%\begin{figure}[t]
%\centering
%\includegraphics[width=8.6cm]{simulation_model_setup.png}
%\caption{Model of the ion trap used in the COMSOL simulations.}
%\label{fig:simulation_model}
%\end{figure}
%%%%%%%%%%
To simulate the potential inside the trap a feature reduced model of the ion trap is used.
%as seen in Figure~\ref{fig:simulation_model}.
The performed simulations follow a similar procedure as already described in~\cite{Noetzold2020:as} and use potentials calculated in the COMSOL Multiphysics 5.4 software in combination with our own ion trajectory molecular dynamics simulation.

Using the AC/DC module from COMSOL with its Electrostatics and Electric Currents submodules the potentials created by the electrodes are calculated utilizing the finite element method. Additionally, COMSOL's infinite element domain feature was employed to model a grounded sphere infinitely far away as a boundary condition. This allows for an overall smaller geometry required to be simulated. The electric field vectors calculated from the electrostatic potential are exported with a grid resolution of $125/\text{mm}$ radially and $15/\text{cm}$ axially. The dynamical electric field vectors due to the wires are only calculated for the center of the trap with a grid resolution of 300/mm. This is done since it is assumed that the axial change of the rf field is negligible and also enables a higher resolution to be used for the grid.

The calculated electric field vectors are exported from COMSOL and in a further step are loaded into a self-developed program for ion trajectory molecular dynamics simulations \cite{Wester2009:jpb}. In the program the ion trajectory of a single ion is calculated by solving Newton’s equations of motion. The radio-frequency field is modulated with a sine function with a frequency of 7.3\,MHz and is superimposed with the constant electric field vectors from the static voltages. The ion is propagated using the well-known Verlet algorithm since it is time efficient to compute and is symplectic, i.\ e.\ it conserves phase space. Because only a discrete amount of electric field vectors are loaded into the program, the force acting on a particle is determined by linear interpolation of the nearby vectors.

The collisions with a helium buffer gas are modelled as random events. After each time step, a random number between zero and one is drawn. A collision occurred if the random number is smaller than $\frac{\Delta t}{\tau} \exp(-\frac{\Delta t}{\tau}) $ with $\Delta t$ being the magnitude of the time step and $\tau$ the average time between collisions. Note that using Poisson's law offers the advantage that the mean number of events is equal to the rate of the the process. In case a collision is detected, the ion and the buffer gas are moved into the center-of-mass frame, with the buffer gas velocity vector components being drawn from 1D Maxwell-Boltzmann distributions. A hard sphere collision is calculated, and the velocity vector components of the ion are changed accordingly.

The time between two collisions with the buffer gas is on average 49\,\si{\micro\second}, the time-step size of the simulation is 3.2\,ns. The position and velocity of the ion is recorded every 1\,ms which corresponds to the time-average of 20 buffer gas collisions. In case the ion is lost from the trap, a new ion is added to the center of the trap and is initialized by waiting for 500 collisions. After 200,000 recordings the simulation is terminated and the temperature is calculated via $T=\pi \frac{m \Bar{v}^{2} }{8k_{\text{B}}}$, where $m$ is the mass of the molecule, $\Bar{v}$ the mean velocity of the particle and $k_{\text{B}}$ the Boltzmann constant. This formula is used to give the notion of a temperature even for the distributions that deviate from the classical Maxwell-Boltzmann distributions.


\begin{thebibliography}{60}%
\makeatletter
\providecommand \@ifxundefined [1]{%
 \@ifx{#1\undefined}
}%
\providecommand \@ifnum [1]{%
 \ifnum #1\expandafter \@firstoftwo
 \else \expandafter \@secondoftwo
 \fi
}%
\providecommand \@ifx [1]{%
 \ifx #1\expandafter \@firstoftwo
 \else \expandafter \@secondoftwo
 \fi
}%
\providecommand \natexlab [1]{#1}%
\providecommand \enquote  [1]{``#1''}%
\providecommand \bibnamefont  [1]{#1}%
\providecommand \bibfnamefont [1]{#1}%
\providecommand \citenamefont [1]{#1}%
\providecommand \href@noop [0]{\@secondoftwo}%
\providecommand \href [0]{\begingroup \@sanitize@url \@href}%
\providecommand \@href[1]{\@@startlink{#1}\@@href}%
\providecommand \@@href[1]{\endgroup#1\@@endlink}%
\providecommand \@sanitize@url [0]{\catcode `\\12\catcode `\$12\catcode
  `\&12\catcode `\#12\catcode `\^12\catcode `\_12\catcode `\%12\relax}%
\providecommand \@@startlink[1]{}%
\providecommand \@@endlink[0]{}%
\providecommand \url  [0]{\begingroup\@sanitize@url \@url }%
\providecommand \@url [1]{\endgroup\@href {#1}{\urlprefix }}%
\providecommand \urlprefix  [0]{URL }%
\providecommand \Eprint [0]{\href }%
\providecommand \doibase [0]{http://dx.doi.org/}%
\providecommand \selectlanguage [0]{\@gobble}%
\providecommand \bibinfo  [0]{\@secondoftwo}%
\providecommand \bibfield  [0]{\@secondoftwo}%
\providecommand \translation [1]{[#1]}%
\providecommand \BibitemOpen [0]{}%
\providecommand \bibitemStop [0]{}%
\providecommand \bibitemNoStop [0]{.\EOS\space}%
\providecommand \EOS [0]{\spacefactor3000\relax}%
\providecommand \BibitemShut  [1]{\csname bibitem#1\endcsname}%
\let\auto@bib@innerbib\@empty
%</preamble>
\bibitem [{\citenamefont {Lykke}\ \emph {et~al.}(1984)\citenamefont {Lykke},
  \citenamefont {Mead},\ and\ \citenamefont {Lineberger}}]{Lykke1984:prl}%
  \BibitemOpen
  \bibfield  {author} {\bibinfo {author} {\bibfnamefont {K.~R.}\ \bibnamefont
  {Lykke}}, \bibinfo {author} {\bibfnamefont {R.~D.}\ \bibnamefont {Mead}}, \
  and\ \bibinfo {author} {\bibfnamefont {W.~C.}\ \bibnamefont {Lineberger}},\
  }\href {\doibase 10.1103/PhysRevLett.52.2221} {\bibfield  {journal} {\bibinfo
   {journal} {Phys. Rev. Lett.}\ }\textbf {\bibinfo {volume} {52}},\ \bibinfo
  {pages} {2221} (\bibinfo {year} {1984})}\BibitemShut {NoStop}%
\bibitem [{\citenamefont {Carelli}\ \emph {et~al.}(2014)\citenamefont
  {Carelli}, \citenamefont {Gianturco}, \citenamefont {Wester},\ and\
  \citenamefont {Satta}}]{Carelli2014:jcp}%
  \BibitemOpen
  \bibfield  {author} {\bibinfo {author} {\bibfnamefont {F.}~\bibnamefont
  {Carelli}}, \bibinfo {author} {\bibfnamefont {F.}~\bibnamefont {Gianturco}},
  \bibinfo {author} {\bibfnamefont {R.}~\bibnamefont {Wester}}, \ and\ \bibinfo
  {author} {\bibfnamefont {M.}~\bibnamefont {Satta}},\ }\href {\doibase
  10.1063/1.4891300} {\bibfield  {journal} {\bibinfo  {journal} {J. Chem.
  Phys.}\ }\textbf {\bibinfo {volume} {141}},\ \bibinfo {pages} {054302}
  (\bibinfo {year} {2014})}\BibitemShut {NoStop}%
\bibitem [{\citenamefont {Yuan}\ \emph {et~al.}(2020)\citenamefont {Yuan},
  \citenamefont {Liu}, \citenamefont {Qian}, \citenamefont {Zhang},
  \citenamefont {Rubenstein},\ and\ \citenamefont {Wang}}]{Yuan2020:prl}%
  \BibitemOpen
  \bibfield  {author} {\bibinfo {author} {\bibfnamefont {D.-F.}\ \bibnamefont
  {Yuan}}, \bibinfo {author} {\bibfnamefont {Y.}~\bibnamefont {Liu}}, \bibinfo
  {author} {\bibfnamefont {C.-H.}\ \bibnamefont {Qian}}, \bibinfo {author}
  {\bibfnamefont {Y.-R.}\ \bibnamefont {Zhang}}, \bibinfo {author}
  {\bibfnamefont {B.~M.}\ \bibnamefont {Rubenstein}}, \ and\ \bibinfo {author}
  {\bibfnamefont {L.-S.}\ \bibnamefont {Wang}},\ }\href {\doibase
  10.1103/physrevlett.125.073003} {\bibfield  {journal} {\bibinfo  {journal}
  {Phys. Rev. Lett.}\ }\textbf {\bibinfo {volume} {125}},\ \bibinfo {pages}
  {073003} (\bibinfo {year} {2020})}\BibitemShut {NoStop}%
\bibitem [{\citenamefont {Simpson}\ \emph {et~al.}(2021)\citenamefont
  {Simpson}, \citenamefont {N\"{o}tzold}, \citenamefont {Michaelsen},
  \citenamefont {Wild}, \citenamefont {Gianturco},\ and\ \citenamefont
  {Wester}}]{Simpson2021:prl}%
  \BibitemOpen
  \bibfield  {author} {\bibinfo {author} {\bibfnamefont {M.}~\bibnamefont
  {Simpson}}, \bibinfo {author} {\bibfnamefont {M.}~\bibnamefont
  {N\"{o}tzold}}, \bibinfo {author} {\bibfnamefont {T.}~\bibnamefont
  {Michaelsen}}, \bibinfo {author} {\bibfnamefont {R.}~\bibnamefont {Wild}},
  \bibinfo {author} {\bibfnamefont {F.~A.}\ \bibnamefont {Gianturco}}, \ and\
  \bibinfo {author} {\bibfnamefont {R.}~\bibnamefont {Wester}},\ }\href
  {\doibase 10.1103/PhysRevLett.127.043001} {\bibfield  {journal} {\bibinfo
  {journal} {Phys. Rev. Lett.}\ }\textbf {\bibinfo {volume} {127}},\ \bibinfo
  {pages} {043001} (\bibinfo {year} {2021})}\BibitemShut {NoStop}%
\bibitem [{\citenamefont {Warring}\ \emph {et~al.}(2009)\citenamefont
  {Warring}, \citenamefont {Amoretti}, \citenamefont {Canali}, \citenamefont
  {Fischer}, \citenamefont {Heyne}, \citenamefont {Meier}, \citenamefont
  {Morhard},\ and\ \citenamefont {Kellerbauer}}]{Warring2009:prl}%
  \BibitemOpen
  \bibfield  {author} {\bibinfo {author} {\bibfnamefont {U.}~\bibnamefont
  {Warring}}, \bibinfo {author} {\bibfnamefont {M.}~\bibnamefont {Amoretti}},
  \bibinfo {author} {\bibfnamefont {C.}~\bibnamefont {Canali}}, \bibinfo
  {author} {\bibfnamefont {A.}~\bibnamefont {Fischer}}, \bibinfo {author}
  {\bibfnamefont {R.}~\bibnamefont {Heyne}}, \bibinfo {author} {\bibfnamefont
  {J.~O.}\ \bibnamefont {Meier}}, \bibinfo {author} {\bibfnamefont
  {C.}~\bibnamefont {Morhard}}, \ and\ \bibinfo {author} {\bibfnamefont
  {A.}~\bibnamefont {Kellerbauer}},\ }\href@noop {} {\bibfield  {journal}
  {\bibinfo  {journal} {Phys. Rev. Lett.}\ }\textbf {\bibinfo {volume} {102}},\
  \bibinfo {pages} {043001} (\bibinfo {year} {2009})}\BibitemShut {NoStop}%
\bibitem [{\citenamefont {Jordan}\ \emph {et~al.}(2015)\citenamefont {Jordan},
  \citenamefont {Cerchiari}, \citenamefont {Fritzsche},\ and\ \citenamefont
  {Kellerbauer}}]{Jordan2015:prl}%
  \BibitemOpen
  \bibfield  {author} {\bibinfo {author} {\bibfnamefont {E.}~\bibnamefont
  {Jordan}}, \bibinfo {author} {\bibfnamefont {G.}~\bibnamefont {Cerchiari}},
  \bibinfo {author} {\bibfnamefont {S.}~\bibnamefont {Fritzsche}}, \ and\
  \bibinfo {author} {\bibfnamefont {A.}~\bibnamefont {Kellerbauer}},\ }\href
  {\doibase 10.1103/physrevlett.115.113001} {\bibfield  {journal} {\bibinfo
  {journal} {Phys. Rev. Lett.}\ }\textbf {\bibinfo {volume} {115}},\ \bibinfo
  {pages} {113001} (\bibinfo {year} {2015})}\BibitemShut {NoStop}%
\bibitem [{\citenamefont {Tang}\ \emph {et~al.}(2019)\citenamefont {Tang},
  \citenamefont {Si}, \citenamefont {Fei}, \citenamefont {Fu}, \citenamefont
  {Lu}, \citenamefont {Brage}, \citenamefont {Liu}, \citenamefont {Chen},\ and\
  \citenamefont {Ning}}]{Tang2019:prl}%
  \BibitemOpen
  \bibfield  {author} {\bibinfo {author} {\bibfnamefont {R.}~\bibnamefont
  {Tang}}, \bibinfo {author} {\bibfnamefont {R.}~\bibnamefont {Si}}, \bibinfo
  {author} {\bibfnamefont {Z.}~\bibnamefont {Fei}}, \bibinfo {author}
  {\bibfnamefont {X.}~\bibnamefont {Fu}}, \bibinfo {author} {\bibfnamefont
  {Y.}~\bibnamefont {Lu}}, \bibinfo {author} {\bibfnamefont {T.}~\bibnamefont
  {Brage}}, \bibinfo {author} {\bibfnamefont {H.}~\bibnamefont {Liu}}, \bibinfo
  {author} {\bibfnamefont {C.}~\bibnamefont {Chen}}, \ and\ \bibinfo {author}
  {\bibfnamefont {C.}~\bibnamefont {Ning}},\ }\href {\doibase
  10.1103/physrevlett.123.203002} {\bibfield  {journal} {\bibinfo  {journal}
  {Phys. Rev. Lett.}\ }\textbf {\bibinfo {volume} {123}},\ \bibinfo {pages}
  {203002} (\bibinfo {year} {2019})}\BibitemShut {NoStop}%
\bibitem [{\citenamefont {Jones}\ \emph {et~al.}(1980)\citenamefont {Jones},
  \citenamefont {Mead}, \citenamefont {Kohler}, \citenamefont {Rosner},\ and\
  \citenamefont {Lineberger}}]{Jones1980:jcp}%
  \BibitemOpen
  \bibfield  {author} {\bibinfo {author} {\bibfnamefont {P.}~\bibnamefont
  {Jones}}, \bibinfo {author} {\bibfnamefont {R.}~\bibnamefont {Mead}},
  \bibinfo {author} {\bibfnamefont {B.}~\bibnamefont {Kohler}}, \bibinfo
  {author} {\bibfnamefont {S.}~\bibnamefont {Rosner}}, \ and\ \bibinfo {author}
  {\bibfnamefont {W.}~\bibnamefont {Lineberger}},\ }\href@noop {} {\bibfield
  {journal} {\bibinfo  {journal} {J. Chem. Phys.}\ }\textbf {\bibinfo {volume}
  {73}},\ \bibinfo {pages} {4419} (\bibinfo {year} {1980})}\BibitemShut
  {NoStop}%
\bibitem [{\citenamefont {Ervin}\ and\ \citenamefont
  {Lineberger}(1991)}]{Ervin1991:jpc}%
  \BibitemOpen
  \bibfield  {author} {\bibinfo {author} {\bibfnamefont {K.~M.}\ \bibnamefont
  {Ervin}}\ and\ \bibinfo {author} {\bibfnamefont {W.}~\bibnamefont
  {Lineberger}},\ }\href {\doibase 10.1021/j100156a026} {\bibfield  {journal}
  {\bibinfo  {journal} {J. Phys. Chem.}\ }\textbf {\bibinfo {volume} {95}},\
  \bibinfo {pages} {1167} (\bibinfo {year} {1991})}\BibitemShut {NoStop}%
\bibitem [{\citenamefont {Shan-Shan}\ \emph {et~al.}(2003)\citenamefont
  {Shan-Shan}, \citenamefont {Xiao-Hua}, \citenamefont {Ben-Xia}, \citenamefont
  {Kakule}, \citenamefont {Sheng-Hai}, \citenamefont {Ying-Chun}, \citenamefont
  {Yu-Yan},\ and\ \citenamefont {Yang-Qin}}]{Shan-Shan2003:cp}%
  \BibitemOpen
  \bibfield  {author} {\bibinfo {author} {\bibfnamefont {Y.}~\bibnamefont
  {Shan-Shan}}, \bibinfo {author} {\bibfnamefont {Y.}~\bibnamefont {Xiao-Hua}},
  \bibinfo {author} {\bibfnamefont {L.}~\bibnamefont {Ben-Xia}}, \bibinfo
  {author} {\bibfnamefont {K.}~\bibnamefont {Kakule}}, \bibinfo {author}
  {\bibfnamefont {W.}~\bibnamefont {Sheng-Hai}}, \bibinfo {author}
  {\bibfnamefont {G.}~\bibnamefont {Ying-Chun}}, \bibinfo {author}
  {\bibfnamefont {L.}~\bibnamefont {Yu-Yan}}, \ and\ \bibinfo {author}
  {\bibfnamefont {C.}~\bibnamefont {Yang-Qin}},\ }\href@noop {} {\bibfield
  {journal} {\bibinfo  {journal} {Chin. Phys.}\ }\textbf {\bibinfo {volume}
  {12}},\ \bibinfo {pages} {745} (\bibinfo {year} {2003})}\BibitemShut
  {NoStop}%
\bibitem [{\citenamefont {Shi}\ \emph {et~al.}(2016)\citenamefont {Shi},
  \citenamefont {Li}, \citenamefont {Meng}, \citenamefont {Wei}, \citenamefont
  {Deng},\ and\ \citenamefont {Yang}}]{Shi2016:ctc}%
  \BibitemOpen
  \bibfield  {author} {\bibinfo {author} {\bibfnamefont {W.}~\bibnamefont
  {Shi}}, \bibinfo {author} {\bibfnamefont {C.}~\bibnamefont {Li}}, \bibinfo
  {author} {\bibfnamefont {H.}~\bibnamefont {Meng}}, \bibinfo {author}
  {\bibfnamefont {J.}~\bibnamefont {Wei}}, \bibinfo {author} {\bibfnamefont
  {L.}~\bibnamefont {Deng}}, \ and\ \bibinfo {author} {\bibfnamefont
  {C.}~\bibnamefont {Yang}},\ }\href {\doibase 10.1016/j.comptc.2016.01.015}
  {\bibfield  {journal} {\bibinfo  {journal} {Comput. Theor. Chem.}\ }\textbf
  {\bibinfo {volume} {1079}},\ \bibinfo {pages} {57} (\bibinfo {year}
  {2016})}\BibitemShut {NoStop}%
\bibitem [{\citenamefont {Yzombard}\ \emph {et~al.}(2015)\citenamefont
  {Yzombard}, \citenamefont {Hamamda}, \citenamefont {Gerber}, \citenamefont
  {Doser},\ and\ \citenamefont {Comparat}}]{Yzombard2015:prl}%
  \BibitemOpen
  \bibfield  {author} {\bibinfo {author} {\bibfnamefont {P.}~\bibnamefont
  {Yzombard}}, \bibinfo {author} {\bibfnamefont {M.}~\bibnamefont {Hamamda}},
  \bibinfo {author} {\bibfnamefont {S.}~\bibnamefont {Gerber}}, \bibinfo
  {author} {\bibfnamefont {M.}~\bibnamefont {Doser}}, \ and\ \bibinfo {author}
  {\bibfnamefont {D.}~\bibnamefont {Comparat}},\ }\href {\doibase
  10.1103/PhysRevLett.114.213001} {\bibfield  {journal} {\bibinfo  {journal}
  {Phys. Rev. Lett.}\ }\textbf {\bibinfo {volume} {114}},\ \bibinfo {pages}
  {213001} (\bibinfo {year} {2015})}\BibitemShut {NoStop}%
\bibitem [{\citenamefont {Gerber}\ \emph {et~al.}(2018)\citenamefont {Gerber},
  \citenamefont {Fesel}, \citenamefont {Doser},\ and\ \citenamefont
  {Comparat}}]{Gerber2018:njp}%
  \BibitemOpen
  \bibfield  {author} {\bibinfo {author} {\bibfnamefont {S.}~\bibnamefont
  {Gerber}}, \bibinfo {author} {\bibfnamefont {J.}~\bibnamefont {Fesel}},
  \bibinfo {author} {\bibfnamefont {M.}~\bibnamefont {Doser}}, \ and\ \bibinfo
  {author} {\bibfnamefont {D.}~\bibnamefont {Comparat}},\ }\href@noop {}
  {\bibfield  {journal} {\bibinfo  {journal} {New J. Phys.}\ }\textbf {\bibinfo
  {volume} {20}},\ \bibinfo {pages} {023024} (\bibinfo {year}
  {2018})}\BibitemShut {NoStop}%
\bibitem [{\citenamefont {Baker}\ \emph {et~al.}(2021)\citenamefont {Baker},
  \citenamefont {Bertsche}, \citenamefont {Capra}, \citenamefont {Carruth},
  \citenamefont {Cesar}, \citenamefont {Charlton}, \citenamefont {Christensen},
  \citenamefont {Collister}, \citenamefont {Mathad}, \citenamefont {Eriksson}
  \emph {et~al.}}]{Baker2021:n}%
  \BibitemOpen
  \bibfield  {author} {\bibinfo {author} {\bibfnamefont {C.}~\bibnamefont
  {Baker}}, \bibinfo {author} {\bibfnamefont {W.}~\bibnamefont {Bertsche}},
  \bibinfo {author} {\bibfnamefont {A.}~\bibnamefont {Capra}}, \bibinfo
  {author} {\bibfnamefont {C.}~\bibnamefont {Carruth}}, \bibinfo {author}
  {\bibfnamefont {C.}~\bibnamefont {Cesar}}, \bibinfo {author} {\bibfnamefont
  {M.}~\bibnamefont {Charlton}}, \bibinfo {author} {\bibfnamefont
  {A.}~\bibnamefont {Christensen}}, \bibinfo {author} {\bibfnamefont
  {R.}~\bibnamefont {Collister}}, \bibinfo {author} {\bibfnamefont {A.~C.}\
  \bibnamefont {Mathad}}, \bibinfo {author} {\bibfnamefont {S.}~\bibnamefont
  {Eriksson}},  \emph {et~al.},\ }\href@noop {} {\bibfield  {journal} {\bibinfo
   {journal} {Nature}\ }\textbf {\bibinfo {volume} {592}},\ \bibinfo {pages}
  {35} (\bibinfo {year} {2021})}\BibitemShut {NoStop}%
\bibitem [{\citenamefont {Barsuhn}(1974)}]{Barsuhn1974:jpb}%
  \BibitemOpen
  \bibfield  {author} {\bibinfo {author} {\bibfnamefont {J.}~\bibnamefont
  {Barsuhn}},\ }\href@noop {} {\bibfield  {journal} {\bibinfo  {journal} {J.
  Phys. B}\ }\textbf {\bibinfo {volume} {7}},\ \bibinfo {pages} {155} (\bibinfo
  {year} {1974})}\BibitemShut {NoStop}%
\bibitem [{\citenamefont {Zeitz}\ \emph {et~al.}(1979)\citenamefont {Zeitz},
  \citenamefont {Peyerimhoff},\ and\ \citenamefont {Buenker}}]{Zeitz1979:cpl}%
  \BibitemOpen
  \bibfield  {author} {\bibinfo {author} {\bibfnamefont {M.}~\bibnamefont
  {Zeitz}}, \bibinfo {author} {\bibfnamefont {S.~D.}\ \bibnamefont
  {Peyerimhoff}}, \ and\ \bibinfo {author} {\bibfnamefont {R.~J.}\ \bibnamefont
  {Buenker}},\ }\href@noop {} {\bibfield  {journal} {\bibinfo  {journal} {Chem.
  Phys. Lett.}\ }\textbf {\bibinfo {volume} {64}},\ \bibinfo {pages} {243}
  (\bibinfo {year} {1979})}\BibitemShut {NoStop}%
\bibitem [{\citenamefont {Dupuis}\ and\ \citenamefont
  {Liu}(1980)}]{Dupuis1980:jcp}%
  \BibitemOpen
  \bibfield  {author} {\bibinfo {author} {\bibfnamefont {M.}~\bibnamefont
  {Dupuis}}\ and\ \bibinfo {author} {\bibfnamefont {B.}~\bibnamefont {Liu}},\
  }\href@noop {} {\bibfield  {journal} {\bibinfo  {journal} {J. Chem. Phys.}\
  }\textbf {\bibinfo {volume} {73}},\ \bibinfo {pages} {337} (\bibinfo {year}
  {1980})}\BibitemShut {NoStop}%
\bibitem [{\citenamefont {Rosmus}\ and\ \citenamefont
  {Werner}(1984)}]{Rosmus1984:jcp}%
  \BibitemOpen
  \bibfield  {author} {\bibinfo {author} {\bibfnamefont {P.}~\bibnamefont
  {Rosmus}}\ and\ \bibinfo {author} {\bibfnamefont {H.-J.}\ \bibnamefont
  {Werner}},\ }\href@noop {} {\bibfield  {journal} {\bibinfo  {journal} {J.
  Chem. Phys.}\ }\textbf {\bibinfo {volume} {80}},\ \bibinfo {pages} {5085}
  (\bibinfo {year} {1984})}\BibitemShut {NoStop}%
\bibitem [{\citenamefont {Nichols}\ and\ \citenamefont
  {Simons}(1987)}]{Nichols1987:jcp}%
  \BibitemOpen
  \bibfield  {author} {\bibinfo {author} {\bibfnamefont {J.~A.}\ \bibnamefont
  {Nichols}}\ and\ \bibinfo {author} {\bibfnamefont {J.}~\bibnamefont
  {Simons}},\ }\href@noop {} {\bibfield  {journal} {\bibinfo  {journal} {J.
  Chem. Phys.}\ }\textbf {\bibinfo {volume} {86}},\ \bibinfo {pages} {6972}
  (\bibinfo {year} {1987})}\BibitemShut {NoStop}%
\bibitem [{\citenamefont {Watts}\ and\ \citenamefont
  {Bartlett}(1992)}]{Watts1992:jcp}%
  \BibitemOpen
  \bibfield  {author} {\bibinfo {author} {\bibfnamefont {J.~D.}\ \bibnamefont
  {Watts}}\ and\ \bibinfo {author} {\bibfnamefont {R.~J.}\ \bibnamefont
  {Bartlett}},\ }\href@noop {} {\bibfield  {journal} {\bibinfo  {journal} {J.
  Chem. Phys.}\ }\textbf {\bibinfo {volume} {96}},\ \bibinfo {pages} {6073}
  (\bibinfo {year} {1992})}\BibitemShut {NoStop}%
\bibitem [{\citenamefont {{\v{S}}edivov\'{a}}\ and\ \citenamefont
  {{\v{S}}pirko}(2006)}]{Sedivova2006:mp}%
  \BibitemOpen
  \bibfield  {author} {\bibinfo {author} {\bibfnamefont {T.}~\bibnamefont
  {{\v{S}}edivov\'{a}}}\ and\ \bibinfo {author} {\bibfnamefont
  {V.}~\bibnamefont {{\v{S}}pirko}},\ }\href {\doibase
  10.1080/00268970600662689} {\bibfield  {journal} {\bibinfo  {journal} {Mol.
  Phys.}\ }\textbf {\bibinfo {volume} {104}},\ \bibinfo {pages} {1999}
  (\bibinfo {year} {2006})}\BibitemShut {NoStop}%
\bibitem [{\citenamefont {Kas}\ \emph {et~al.}(2019)\citenamefont {Kas},
  \citenamefont {Loreau}, \citenamefont {Li\'{e}vin},\ and\ \citenamefont
  {Vaeck}}]{Kas2019:pra}%
  \BibitemOpen
  \bibfield  {author} {\bibinfo {author} {\bibfnamefont {M.}~\bibnamefont
  {Kas}}, \bibinfo {author} {\bibfnamefont {J.}~\bibnamefont {Loreau}},
  \bibinfo {author} {\bibfnamefont {J.}~\bibnamefont {Li\'{e}vin}}, \ and\
  \bibinfo {author} {\bibfnamefont {N.}~\bibnamefont {Vaeck}},\ }\href@noop {}
  {\bibfield  {journal} {\bibinfo  {journal} {Phys. Rev. A}\ }\textbf {\bibinfo
  {volume} {99}},\ \bibinfo {pages} {042702} (\bibinfo {year}
  {2019})}\BibitemShut {NoStop}%
\bibitem [{\citenamefont {Gulania}\ \emph {et~al.}(2019)\citenamefont
  {Gulania}, \citenamefont {Jagau},\ and\ \citenamefont
  {Krylov}}]{Gulania2019:fd}%
  \BibitemOpen
  \bibfield  {author} {\bibinfo {author} {\bibfnamefont {S.}~\bibnamefont
  {Gulania}}, \bibinfo {author} {\bibfnamefont {T.-C.}\ \bibnamefont {Jagau}},
  \ and\ \bibinfo {author} {\bibfnamefont {A.~I.}\ \bibnamefont {Krylov}},\
  }\href@noop {} {\bibfield  {journal} {\bibinfo  {journal} {Faraday Discuss.}\
  }\textbf {\bibinfo {volume} {217}},\ \bibinfo {pages} {514} (\bibinfo {year}
  {2019})}\BibitemShut {NoStop}%
\bibitem [{\citenamefont {Herzberg}\ and\ \citenamefont
  {Lagerqvist}(1968)}]{Herzberg1968:cjp}%
  \BibitemOpen
  \bibfield  {author} {\bibinfo {author} {\bibfnamefont {G.}~\bibnamefont
  {Herzberg}}\ and\ \bibinfo {author} {\bibfnamefont {A.}~\bibnamefont
  {Lagerqvist}},\ }\href@noop {} {\bibfield  {journal} {\bibinfo  {journal}
  {Can. J. Phys.}\ }\textbf {\bibinfo {volume} {46}},\ \bibinfo {pages} {2363}
  (\bibinfo {year} {1968})}\BibitemShut {NoStop}%
\bibitem [{\citenamefont {Mead}\ \emph {et~al.}(1985)\citenamefont {Mead},
  \citenamefont {Hefter}, \citenamefont {Schulz},\ and\ \citenamefont
  {Lineberger}}]{Mead1985:jcp}%
  \BibitemOpen
  \bibfield  {author} {\bibinfo {author} {\bibfnamefont {R.~D.}\ \bibnamefont
  {Mead}}, \bibinfo {author} {\bibfnamefont {U.}~\bibnamefont {Hefter}},
  \bibinfo {author} {\bibfnamefont {P.}~\bibnamefont {Schulz}}, \ and\ \bibinfo
  {author} {\bibfnamefont {W.}~\bibnamefont {Lineberger}},\ }\href@noop {}
  {\bibfield  {journal} {\bibinfo  {journal} {J. Chem. Phys.}\ }\textbf
  {\bibinfo {volume} {82}},\ \bibinfo {pages} {1723} (\bibinfo {year}
  {1985})}\BibitemShut {NoStop}%
\bibitem [{\citenamefont {Rehfuss}\ \emph {et~al.}(1988)\citenamefont
  {Rehfuss}, \citenamefont {Liu}, \citenamefont {Dinelli}, \citenamefont
  {Jagod}, \citenamefont {Ho}, \citenamefont {Crofton},\ and\ \citenamefont
  {Oka}}]{Rehfuss1988:jcp}%
  \BibitemOpen
  \bibfield  {author} {\bibinfo {author} {\bibfnamefont {B.~D.}\ \bibnamefont
  {Rehfuss}}, \bibinfo {author} {\bibfnamefont {D.-J.}\ \bibnamefont {Liu}},
  \bibinfo {author} {\bibfnamefont {B.~M.}\ \bibnamefont {Dinelli}}, \bibinfo
  {author} {\bibfnamefont {M.-F.}\ \bibnamefont {Jagod}}, \bibinfo {author}
  {\bibfnamefont {W.~C.}\ \bibnamefont {Ho}}, \bibinfo {author} {\bibfnamefont
  {M.~W.}\ \bibnamefont {Crofton}}, \ and\ \bibinfo {author} {\bibfnamefont
  {T.}~\bibnamefont {Oka}},\ }\href@noop {} {\bibfield  {journal} {\bibinfo
  {journal} {J. Chem. Phys.}\ }\textbf {\bibinfo {volume} {89}},\ \bibinfo
  {pages} {129} (\bibinfo {year} {1988})}\BibitemShut {NoStop}%
\bibitem [{\citenamefont {Lineberger}\ and\ \citenamefont
  {Patterson}(1972)}]{Lineberger1972:cpl}%
  \BibitemOpen
  \bibfield  {author} {\bibinfo {author} {\bibfnamefont {W.}~\bibnamefont
  {Lineberger}}\ and\ \bibinfo {author} {\bibfnamefont {T.}~\bibnamefont
  {Patterson}},\ }\href@noop {} {\bibfield  {journal} {\bibinfo  {journal}
  {Chem. Phys. Lett.}\ }\textbf {\bibinfo {volume} {13}},\ \bibinfo {pages}
  {40} (\bibinfo {year} {1972})}\BibitemShut {NoStop}%
\bibitem [{\citenamefont {Royen}\ and\ \citenamefont
  {Zackrisson}(1992)}]{Royen1992:jms}%
  \BibitemOpen
  \bibfield  {author} {\bibinfo {author} {\bibfnamefont {P.}~\bibnamefont
  {Royen}}\ and\ \bibinfo {author} {\bibfnamefont {M.}~\bibnamefont
  {Zackrisson}},\ }\href@noop {} {\bibfield  {journal} {\bibinfo  {journal} {J.
  Mol. Spectrosc.}\ }\textbf {\bibinfo {volume} {155}},\ \bibinfo {pages} {427}
  (\bibinfo {year} {1992})}\BibitemShut {NoStop}%
\bibitem [{\citenamefont {de~Beer}\ \emph {et~al.}(1995)\citenamefont
  {de~Beer}, \citenamefont {Kim}, \citenamefont {Neumark}, \citenamefont
  {Gunion},\ and\ \citenamefont {Lineberger}}]{Beer1995:jpc}%
  \BibitemOpen
  \bibfield  {author} {\bibinfo {author} {\bibfnamefont {E.}~\bibnamefont
  {de~Beer}}, \bibinfo {author} {\bibfnamefont {E.~H.}\ \bibnamefont {Kim}},
  \bibinfo {author} {\bibfnamefont {D.~M.}\ \bibnamefont {Neumark}}, \bibinfo
  {author} {\bibfnamefont {R.~F.}\ \bibnamefont {Gunion}}, \ and\ \bibinfo
  {author} {\bibfnamefont {W.~C.}\ \bibnamefont {Lineberger}},\ }\href
  {\doibase 10.1021/j100037a009} {\bibfield  {journal} {\bibinfo  {journal} {J.
  Phys. Chem.}\ }\textbf {\bibinfo {volume} {99}},\ \bibinfo {pages} {13627}
  (\bibinfo {year} {1995})}\BibitemShut {NoStop}%
\bibitem [{\citenamefont {Bragg}\ \emph {et~al.}(2003)\citenamefont {Bragg},
  \citenamefont {Wester}, \citenamefont {Davis}, \citenamefont {Kammrath},\
  and\ \citenamefont {Neumark}}]{Bragg2003:cpl}%
  \BibitemOpen
  \bibfield  {author} {\bibinfo {author} {\bibfnamefont {A.~E.}\ \bibnamefont
  {Bragg}}, \bibinfo {author} {\bibfnamefont {R.}~\bibnamefont {Wester}},
  \bibinfo {author} {\bibfnamefont {A.~V.}\ \bibnamefont {Davis}}, \bibinfo
  {author} {\bibfnamefont {A.}~\bibnamefont {Kammrath}}, \ and\ \bibinfo
  {author} {\bibfnamefont {D.~M.}\ \bibnamefont {Neumark}},\ }\href@noop {}
  {\bibfield  {journal} {\bibinfo  {journal} {Chem. Phys. Lett.}\ }\textbf
  {\bibinfo {volume} {376}},\ \bibinfo {pages} {767} (\bibinfo {year}
  {2003})}\BibitemShut {NoStop}%
\bibitem [{\citenamefont {Nakajima}(2017)}]{Nakajima2017:jms}%
  \BibitemOpen
  \bibfield  {author} {\bibinfo {author} {\bibfnamefont {M.}~\bibnamefont
  {Nakajima}},\ }\href@noop {} {\bibfield  {journal} {\bibinfo  {journal} {J.
  Mol. Spectrosc.}\ }\textbf {\bibinfo {volume} {331}},\ \bibinfo {pages} {106}
  (\bibinfo {year} {2017})}\BibitemShut {NoStop}%
\bibitem [{\citenamefont {Pedersen}\ \emph {et~al.}(1998)\citenamefont
  {Pedersen}, \citenamefont {Brink}, \citenamefont {Andersen}, \citenamefont
  {Bjerre}, \citenamefont {Hvelplund}, \citenamefont {Kella},\ and\
  \citenamefont {Shen}}]{Pedersen1998:jcp}%
  \BibitemOpen
  \bibfield  {author} {\bibinfo {author} {\bibfnamefont {H.}~\bibnamefont
  {Pedersen}}, \bibinfo {author} {\bibfnamefont {C.}~\bibnamefont {Brink}},
  \bibinfo {author} {\bibfnamefont {L.}~\bibnamefont {Andersen}}, \bibinfo
  {author} {\bibfnamefont {N.}~\bibnamefont {Bjerre}}, \bibinfo {author}
  {\bibfnamefont {P.}~\bibnamefont {Hvelplund}}, \bibinfo {author}
  {\bibfnamefont {D.}~\bibnamefont {Kella}}, \ and\ \bibinfo {author}
  {\bibfnamefont {H.}~\bibnamefont {Shen}},\ }\href@noop {} {\bibfield
  {journal} {\bibinfo  {journal} {J. Chem. Phys.}\ }\textbf {\bibinfo {volume}
  {109}},\ \bibinfo {pages} {5849} (\bibinfo {year} {1998})}\BibitemShut
  {NoStop}%
\bibitem [{\citenamefont {Iida}\ \emph {et~al.}(2020)\citenamefont {Iida},
  \citenamefont {Kuma}, \citenamefont {Tanuma}, \citenamefont {Azuma},\ and\
  \citenamefont {Shiromaru}}]{Iida2020:jpcl}%
  \BibitemOpen
  \bibfield  {author} {\bibinfo {author} {\bibfnamefont {S.}~\bibnamefont
  {Iida}}, \bibinfo {author} {\bibfnamefont {S.}~\bibnamefont {Kuma}}, \bibinfo
  {author} {\bibfnamefont {H.}~\bibnamefont {Tanuma}}, \bibinfo {author}
  {\bibfnamefont {T.}~\bibnamefont {Azuma}}, \ and\ \bibinfo {author}
  {\bibfnamefont {H.}~\bibnamefont {Shiromaru}},\ }\href@noop {} {\bibfield
  {journal} {\bibinfo  {journal} {J. Phys. Chem. Lett.}\ }\textbf {\bibinfo
  {volume} {11}},\ \bibinfo {pages} {10526} (\bibinfo {year}
  {2020})}\BibitemShut {NoStop}%
\bibitem [{\citenamefont {Perry-Sassmannshausen}\ \emph
  {et~al.}(2020)\citenamefont {Perry-Sassmannshausen}, \citenamefont {Buhr},
  \citenamefont {Borovik}, \citenamefont {Martins}, \citenamefont {Reinwardt},
  \citenamefont {Ricz}, \citenamefont {Stock}, \citenamefont {Trinter},
  \citenamefont {M\"uller}, \citenamefont {Fritzsche},\ and\ \citenamefont
  {Schippers}}]{Perry-Sassmannshausen2020:prl}%
  \BibitemOpen
  \bibfield  {author} {\bibinfo {author} {\bibfnamefont {A.}~\bibnamefont
  {Perry-Sassmannshausen}}, \bibinfo {author} {\bibfnamefont {T.}~\bibnamefont
  {Buhr}}, \bibinfo {author} {\bibfnamefont {A.}~\bibnamefont {Borovik}},
  \bibinfo {author} {\bibfnamefont {M.}~\bibnamefont {Martins}}, \bibinfo
  {author} {\bibfnamefont {S.}~\bibnamefont {Reinwardt}}, \bibinfo {author}
  {\bibfnamefont {S.}~\bibnamefont {Ricz}}, \bibinfo {author} {\bibfnamefont
  {S.~O.}\ \bibnamefont {Stock}}, \bibinfo {author} {\bibfnamefont
  {F.}~\bibnamefont {Trinter}}, \bibinfo {author} {\bibfnamefont
  {A.}~\bibnamefont {M\"uller}}, \bibinfo {author} {\bibfnamefont
  {S.}~\bibnamefont {Fritzsche}}, \ and\ \bibinfo {author} {\bibfnamefont
  {S.}~\bibnamefont {Schippers}},\ }\href {\doibase
  10.1103/PhysRevLett.124.083203} {\bibfield  {journal} {\bibinfo  {journal}
  {Phys. Rev. Lett.}\ }\textbf {\bibinfo {volume} {124}},\ \bibinfo {pages}
  {083203} (\bibinfo {year} {2020})}\BibitemShut {NoStop}%
\bibitem [{\citenamefont {Germann}\ \emph {et~al.}(2014)\citenamefont
  {Germann}, \citenamefont {Tong},\ and\ \citenamefont
  {Willitsch}}]{Germann2014:np}%
  \BibitemOpen
  \bibfield  {author} {\bibinfo {author} {\bibfnamefont {M.}~\bibnamefont
  {Germann}}, \bibinfo {author} {\bibfnamefont {X.}~\bibnamefont {Tong}}, \
  and\ \bibinfo {author} {\bibfnamefont {S.}~\bibnamefont {Willitsch}},\ }\href
  {\doibase 10.1038/nphys3085} {\bibfield  {journal} {\bibinfo  {journal} {Nat.
  Phys.}\ }\textbf {\bibinfo {volume} {10}},\ \bibinfo {pages} {820} (\bibinfo
  {year} {2014})}\BibitemShut {NoStop}%
\bibitem [{\citenamefont {Lambert}\ \emph {et~al.}(1995)\citenamefont
  {Lambert}, \citenamefont {Sheffer},\ and\ \citenamefont
  {Federman}}]{Lambert1995:aj}%
  \BibitemOpen
  \bibfield  {author} {\bibinfo {author} {\bibfnamefont {D.~L.}\ \bibnamefont
  {Lambert}}, \bibinfo {author} {\bibfnamefont {Y.}~\bibnamefont {Sheffer}}, \
  and\ \bibinfo {author} {\bibfnamefont {S.}~\bibnamefont {Federman}},\
  }\href@noop {} {\bibfield  {journal} {\bibinfo  {journal} {Astrophys. J.}\
  }\textbf {\bibinfo {volume} {438}},\ \bibinfo {pages} {740} (\bibinfo {year}
  {1995})}\BibitemShut {NoStop}%
\bibitem [{\citenamefont {Souza}\ and\ \citenamefont
  {LUTz}(1977)}]{Souza1977:aj}%
  \BibitemOpen
  \bibfield  {author} {\bibinfo {author} {\bibfnamefont {S.~P.}\ \bibnamefont
  {Souza}}\ and\ \bibinfo {author} {\bibfnamefont {B.~L.}\ \bibnamefont
  {LUTz}},\ }\href@noop {} {\bibfield  {journal} {\bibinfo  {journal}
  {Astrophys. J.}\ }\textbf {\bibinfo {volume} {216}},\ \bibinfo {pages} {L49}
  (\bibinfo {year} {1977})}\BibitemShut {NoStop}%
\bibitem [{\citenamefont {Lambert}\ \emph {et~al.}(1986)\citenamefont
  {Lambert}, \citenamefont {Gustafsson}, \citenamefont {Eriksson},\ and\
  \citenamefont {Hinkle}}]{Lambert1986:ajss}%
  \BibitemOpen
  \bibfield  {author} {\bibinfo {author} {\bibfnamefont {D.~L.}\ \bibnamefont
  {Lambert}}, \bibinfo {author} {\bibfnamefont {B.}~\bibnamefont {Gustafsson}},
  \bibinfo {author} {\bibfnamefont {K.}~\bibnamefont {Eriksson}}, \ and\
  \bibinfo {author} {\bibfnamefont {K.~H.}\ \bibnamefont {Hinkle}},\
  }\href@noop {} {\bibfield  {journal} {\bibinfo  {journal} {Astrophys. J.
  Suppl. Ser.}\ }\textbf {\bibinfo {volume} {62}},\ \bibinfo {pages} {373}
  (\bibinfo {year} {1986})}\BibitemShut {NoStop}%
\bibitem [{\citenamefont {Civi{\v{s}}}\ \emph {et~al.}(2005)\citenamefont
  {Civi{\v{s}}}, \citenamefont {Hosaki}, \citenamefont {Kagi}, \citenamefont
  {Izumiura}, \citenamefont {Yanagisawa}, \citenamefont {{\v{S}}edivcov\'{a}},\
  and\ \citenamefont {Kawaguchi}}]{Civis2005:pasj}%
  \BibitemOpen
  \bibfield  {author} {\bibinfo {author} {\bibfnamefont {S.}~\bibnamefont
  {Civi{\v{s}}}}, \bibinfo {author} {\bibfnamefont {Y.}~\bibnamefont {Hosaki}},
  \bibinfo {author} {\bibfnamefont {E.}~\bibnamefont {Kagi}}, \bibinfo {author}
  {\bibfnamefont {H.}~\bibnamefont {Izumiura}}, \bibinfo {author}
  {\bibfnamefont {K.}~\bibnamefont {Yanagisawa}}, \bibinfo {author}
  {\bibfnamefont {T.}~\bibnamefont {{\v{S}}edivcov\'{a}}}, \ and\ \bibinfo
  {author} {\bibfnamefont {K.}~\bibnamefont {Kawaguchi}},\ }\href@noop {}
  {\bibfield  {journal} {\bibinfo  {journal} {Publ. Astron. Soc. Jpn.}\
  }\textbf {\bibinfo {volume} {57}},\ \bibinfo {pages} {605} (\bibinfo {year}
  {2005})}\BibitemShut {NoStop}%
\bibitem [{\citenamefont {Leutwyler}\ \emph {et~al.}(1982)\citenamefont
  {Leutwyler}, \citenamefont {Maier},\ and\ \citenamefont
  {Misev}}]{Leutwyler1982:cpl}%
  \BibitemOpen
  \bibfield  {author} {\bibinfo {author} {\bibfnamefont {S.}~\bibnamefont
  {Leutwyler}}, \bibinfo {author} {\bibfnamefont {J.~P.}\ \bibnamefont
  {Maier}}, \ and\ \bibinfo {author} {\bibfnamefont {L.}~\bibnamefont
  {Misev}},\ }\href {\doibase 10.1016/0009-2614(82)83642-7} {\bibfield
  {journal} {\bibinfo  {journal} {Chern. Phys. Lett.}\ }\textbf {\bibinfo
  {volume} {91}},\ \bibinfo {pages} {206} (\bibinfo {year} {1982})}\BibitemShut
  {NoStop}%
\bibitem [{\citenamefont {Geistlinger}\ \emph {et~al.}(2021)\citenamefont
  {Geistlinger}, \citenamefont {Fischer}, \citenamefont {Spieler},
  \citenamefont {Remmers}, \citenamefont {Duensing}, \citenamefont {Dahlmann},
  \citenamefont {Endres},\ and\ \citenamefont {Wester}}]{Geistlinger2021:rsi}%
  \BibitemOpen
  \bibfield  {author} {\bibinfo {author} {\bibfnamefont {K.}~\bibnamefont
  {Geistlinger}}, \bibinfo {author} {\bibfnamefont {M.}~\bibnamefont
  {Fischer}}, \bibinfo {author} {\bibfnamefont {S.}~\bibnamefont {Spieler}},
  \bibinfo {author} {\bibfnamefont {L.}~\bibnamefont {Remmers}}, \bibinfo
  {author} {\bibfnamefont {F.}~\bibnamefont {Duensing}}, \bibinfo {author}
  {\bibfnamefont {F.}~\bibnamefont {Dahlmann}}, \bibinfo {author}
  {\bibfnamefont {E.}~\bibnamefont {Endres}}, \ and\ \bibinfo {author}
  {\bibfnamefont {R.}~\bibnamefont {Wester}},\ }\href {\doibase
  10.1063/5.0040866} {\bibfield  {journal} {\bibinfo  {journal} {Rev. Sci.
  Instrum.}\ }\textbf {\bibinfo {volume} {92}},\ \bibinfo {pages} {023204}
  (\bibinfo {year} {2021})}\BibitemShut {NoStop}%
\bibitem [{\citenamefont {Best}\ \emph {et~al.}(2011)\citenamefont {Best},
  \citenamefont {Otto}, \citenamefont {Trippel}, \citenamefont {Hlavenka},
  \citenamefont {von Zastrow}, \citenamefont {Eisenbach}, \citenamefont
  {Jezouin}, \citenamefont {Wester}, \citenamefont {Vigren}, \citenamefont
  {Hamberg},\ and\ \citenamefont {Geppert}}]{Best2011:aj}%
  \BibitemOpen
  \bibfield  {author} {\bibinfo {author} {\bibfnamefont {T.}~\bibnamefont
  {Best}}, \bibinfo {author} {\bibfnamefont {R.}~\bibnamefont {Otto}}, \bibinfo
  {author} {\bibfnamefont {S.}~\bibnamefont {Trippel}}, \bibinfo {author}
  {\bibfnamefont {P.}~\bibnamefont {Hlavenka}}, \bibinfo {author}
  {\bibfnamefont {A.}~\bibnamefont {von Zastrow}}, \bibinfo {author}
  {\bibfnamefont {S.}~\bibnamefont {Eisenbach}}, \bibinfo {author}
  {\bibfnamefont {S.}~\bibnamefont {Jezouin}}, \bibinfo {author} {\bibfnamefont
  {R.}~\bibnamefont {Wester}}, \bibinfo {author} {\bibfnamefont
  {E.}~\bibnamefont {Vigren}}, \bibinfo {author} {\bibfnamefont
  {M.}~\bibnamefont {Hamberg}}, \ and\ \bibinfo {author} {\bibfnamefont
  {W.~D.}\ \bibnamefont {Geppert}},\ }\href
  {http://stacks.iop.org/0004-637X/742/i=2/a=63} {\bibfield  {journal}
  {\bibinfo  {journal} {Astrophys. J.}\ }\textbf {\bibinfo {volume} {742}},\
  \bibinfo {pages} {63} (\bibinfo {year} {2011})}\BibitemShut {NoStop}%
\bibitem [{\citenamefont {Mant}\ \emph
  {et~al.}(2020{\natexlab{a}})\citenamefont {Mant}, \citenamefont {Gianturco},
  \citenamefont {Wester}, \citenamefont {Gonzalez},\ and\ \citenamefont
  {Yurtsever}}]{Mant2020:ijms}%
  \BibitemOpen
  \bibfield  {author} {\bibinfo {author} {\bibfnamefont {B.~P.}\ \bibnamefont
  {Mant}}, \bibinfo {author} {\bibfnamefont {F.~A.}\ \bibnamefont {Gianturco}},
  \bibinfo {author} {\bibfnamefont {R.}~\bibnamefont {Wester}}, \bibinfo
  {author} {\bibfnamefont {L.}~\bibnamefont {Gonzalez}}, \ and\ \bibinfo
  {author} {\bibfnamefont {E.}~\bibnamefont {Yurtsever}},\ }\href {\doibase
  10.1016/j.ijms.2020.116426} {\bibfield  {journal} {\bibinfo  {journal} {Int.
  J. Mass. Spectrom.}\ }\textbf {\bibinfo {volume} {457}},\ \bibinfo {pages}
  {116426} (\bibinfo {year} {2020}{\natexlab{a}})}\BibitemShut {NoStop}%
\bibitem [{\citenamefont {Mant}\ \emph
  {et~al.}(2020{\natexlab{b}})\citenamefont {Mant}, \citenamefont {Gianturco},
  \citenamefont {Wester}, \citenamefont {Yurtsever},\ and\ \citenamefont
  {Gonz\'{a}lez-S\'{a}nchez}}]{Mant2020:pra}%
  \BibitemOpen
  \bibfield  {author} {\bibinfo {author} {\bibfnamefont {B.~P.}\ \bibnamefont
  {Mant}}, \bibinfo {author} {\bibfnamefont {F.~A.}\ \bibnamefont {Gianturco}},
  \bibinfo {author} {\bibfnamefont {R.}~\bibnamefont {Wester}}, \bibinfo
  {author} {\bibfnamefont {E.}~\bibnamefont {Yurtsever}}, \ and\ \bibinfo
  {author} {\bibfnamefont {L.}~\bibnamefont {Gonz\'{a}lez-S\'{a}nchez}},\
  }\href {\doibase 10.1103/PhysRevA.102.062810} {\bibfield  {journal} {\bibinfo
   {journal} {Phys. Rev. A}\ }\textbf {\bibinfo {volume} {102}},\ \bibinfo
  {pages} {062810} (\bibinfo {year} {2020}{\natexlab{b}})}\BibitemShut
  {NoStop}%
\bibitem [{\citenamefont {Herzberg}\ and\ \citenamefont
  {Huber}(1950)}]{Herzberg1950}%
  \BibitemOpen
  \bibfield  {author} {\bibinfo {author} {\bibfnamefont {G.}~\bibnamefont
  {Herzberg}}\ and\ \bibinfo {author} {\bibfnamefont {K.~P.}\ \bibnamefont
  {Huber}},\ }\enquote {\bibinfo {title} {Molecular spectra and molecular
  structure: Spectra of diatomic moleculesa},}\ \ (\bibinfo  {publisher} {Van
  Nostrand},\ \bibinfo {year} {1950})\ pp.\ \bibinfo {pages}
  {247--251}\BibitemShut {NoStop}%
\bibitem [{\citenamefont {Asvany}\ and\ \citenamefont
  {Schlemmer}(2009)}]{Asvany2009:ijms}%
  \BibitemOpen
  \bibfield  {author} {\bibinfo {author} {\bibfnamefont {O.}~\bibnamefont
  {Asvany}}\ and\ \bibinfo {author} {\bibfnamefont {S.}~\bibnamefont
  {Schlemmer}},\ }\href@noop {} {\bibfield  {journal} {\bibinfo  {journal}
  {Int. J. Mass Spectrom.}\ }\textbf {\bibinfo {volume} {279}},\ \bibinfo
  {pages} {147} (\bibinfo {year} {2009})}\BibitemShut {NoStop}%
\bibitem [{\citenamefont {N\"{o}tzold}\ \emph {et~al.}(2020)\citenamefont
  {N\"{o}tzold}, \citenamefont {Hassan}, \citenamefont {Tauch}, \citenamefont
  {Endres}, \citenamefont {Wester},\ and\ \citenamefont
  {Weidem\"{u}ller}}]{Noetzold2020:as}%
  \BibitemOpen
  \bibfield  {author} {\bibinfo {author} {\bibfnamefont {M.}~\bibnamefont
  {N\"{o}tzold}}, \bibinfo {author} {\bibfnamefont {S.~Z.}\ \bibnamefont
  {Hassan}}, \bibinfo {author} {\bibfnamefont {J.}~\bibnamefont {Tauch}},
  \bibinfo {author} {\bibfnamefont {E.}~\bibnamefont {Endres}}, \bibinfo
  {author} {\bibfnamefont {R.}~\bibnamefont {Wester}}, \ and\ \bibinfo {author}
  {\bibfnamefont {M.}~\bibnamefont {Weidem\"{u}ller}},\ }\href {\doibase
  10.3390/app10155264} {\bibfield  {journal} {\bibinfo  {journal} {Appl. Sci.}\
  }\textbf {\bibinfo {volume} {10}},\ \bibinfo {pages} {5264} (\bibinfo {year}
  {2020})}\BibitemShut {NoStop}%
\bibitem [{\citenamefont {H\"{o}ltkemeier}\ \emph {et~al.}(2016)\citenamefont
  {H\"{o}ltkemeier}, \citenamefont {Weckesser}, \citenamefont
  {L\'{o}pez-Carrera},\ and\ \citenamefont
  {Weidem\"{u}ller}}]{Hoeltkemeier2016:prl}%
  \BibitemOpen
  \bibfield  {author} {\bibinfo {author} {\bibfnamefont {B.}~\bibnamefont
  {H\"{o}ltkemeier}}, \bibinfo {author} {\bibfnamefont {P.}~\bibnamefont
  {Weckesser}}, \bibinfo {author} {\bibfnamefont {H.}~\bibnamefont
  {L\'{o}pez-Carrera}}, \ and\ \bibinfo {author} {\bibfnamefont
  {M.}~\bibnamefont {Weidem\"{u}ller}},\ }\href {\doibase
  10.1103/PhysRevLett.116.233003} {\bibfield  {journal} {\bibinfo  {journal}
  {Phys. Rev. Lett.}\ }\textbf {\bibinfo {volume} {116}},\ \bibinfo {pages}
  {233003} (\bibinfo {year} {2016})}\BibitemShut {NoStop}%
\bibitem [{\citenamefont {Jusko}\ \emph {et~al.}(2014)\citenamefont {Jusko},
  \citenamefont {Asvany}, \citenamefont {Wallerstein}, \citenamefont
  {Br\"{u}nken},\ and\ \citenamefont {Schlemmer}}]{Jusko2014:prl}%
  \BibitemOpen
  \bibfield  {author} {\bibinfo {author} {\bibfnamefont {P.}~\bibnamefont
  {Jusko}}, \bibinfo {author} {\bibfnamefont {O.}~\bibnamefont {Asvany}},
  \bibinfo {author} {\bibfnamefont {A.-C.}\ \bibnamefont {Wallerstein}},
  \bibinfo {author} {\bibfnamefont {S.}~\bibnamefont {Br\"{u}nken}}, \ and\
  \bibinfo {author} {\bibfnamefont {S.}~\bibnamefont {Schlemmer}},\ }\href
  {\doibase 10.1103/PhysRevLett.112.253005} {\bibfield  {journal} {\bibinfo
  {journal} {Phys. Rev. Lett.}\ }\textbf {\bibinfo {volume} {112}},\ \bibinfo
  {pages} {253005} (\bibinfo {year} {2014})}\BibitemShut {NoStop}%
\bibitem [{\citenamefont {Dom{\'{e}}nech}\ \emph
  {et~al.}(2018{\natexlab{a}})\citenamefont {Dom{\'{e}}nech}, \citenamefont
  {Jusko}, \citenamefont {Schlemmer},\ and\ \citenamefont
  {Asvany}}]{Domenech2018a:aj}%
  \BibitemOpen
  \bibfield  {author} {\bibinfo {author} {\bibfnamefont {J.~L.}\ \bibnamefont
  {Dom{\'{e}}nech}}, \bibinfo {author} {\bibfnamefont {P.}~\bibnamefont
  {Jusko}}, \bibinfo {author} {\bibfnamefont {S.}~\bibnamefont {Schlemmer}}, \
  and\ \bibinfo {author} {\bibfnamefont {O.}~\bibnamefont {Asvany}},\ }\href
  {\doibase 10.3847/1538-4357/aab36a} {\bibfield  {journal} {\bibinfo
  {journal} {Astrophys. J.}\ }\textbf {\bibinfo {volume} {857}},\ \bibinfo
  {pages} {61} (\bibinfo {year} {2018}{\natexlab{a}})}\BibitemShut {NoStop}%
\bibitem [{\citenamefont {Dom{\'{e}}nech}\ \emph
  {et~al.}(2018{\natexlab{b}})\citenamefont {Dom{\'{e}}nech}, \citenamefont
  {Schlemmer},\ and\ \citenamefont {Asvany}}]{Domenech2018b:aj}%
  \BibitemOpen
  \bibfield  {author} {\bibinfo {author} {\bibfnamefont {J.~L.}\ \bibnamefont
  {Dom{\'{e}}nech}}, \bibinfo {author} {\bibfnamefont {S.}~\bibnamefont
  {Schlemmer}}, \ and\ \bibinfo {author} {\bibfnamefont {O.}~\bibnamefont
  {Asvany}},\ }\href {\doibase 10.3847/1538-4357/aadf83} {\bibfield  {journal}
  {\bibinfo  {journal} {Astrophys. J.}\ }\textbf {\bibinfo {volume} {866}},\
  \bibinfo {pages} {158} (\bibinfo {year} {2018}{\natexlab{b}})}\BibitemShut
  {NoStop}%
\bibitem [{\citenamefont {Lakhmanskaya}\ \emph {et~al.}(2020)\citenamefont
  {Lakhmanskaya}, \citenamefont {Simpson},\ and\ \citenamefont
  {Wester}}]{Lakhmanskaya2020:pra}%
  \BibitemOpen
  \bibfield  {author} {\bibinfo {author} {\bibfnamefont {O.}~\bibnamefont
  {Lakhmanskaya}}, \bibinfo {author} {\bibfnamefont {M.}~\bibnamefont
  {Simpson}}, \ and\ \bibinfo {author} {\bibfnamefont {R.}~\bibnamefont
  {Wester}},\ }\href {\doibase 10.1103/physreva.102.012809} {\bibfield
  {journal} {\bibinfo  {journal} {Phys. Rev. A}\ }\textbf {\bibinfo {volume}
  {102}},\ \bibinfo {pages} {012809} (\bibinfo {year} {2020})}\BibitemShut
  {NoStop}%
\bibitem [{\citenamefont {Rouse}\ and\ \citenamefont
  {Willitsch}(2017)}]{Rouse2017:prl}%
  \BibitemOpen
  \bibfield  {author} {\bibinfo {author} {\bibfnamefont {I.}~\bibnamefont
  {Rouse}}\ and\ \bibinfo {author} {\bibfnamefont {S.}~\bibnamefont
  {Willitsch}},\ }\href {\doibase 10.1103/PhysRevLett.118.143401} {\bibfield
  {journal} {\bibinfo  {journal} {Phys. Rev. Lett.}\ }\textbf {\bibinfo
  {volume} {118}},\ \bibinfo {pages} {143401} (\bibinfo {year}
  {2017})}\BibitemShut {NoStop}%
\bibitem [{\citenamefont {Otto}\ \emph {et~al.}(2009)\citenamefont {Otto},
  \citenamefont {Hlavenka}, \citenamefont {Trippel}, \citenamefont {Mikosch},
  \citenamefont {Singer}, \citenamefont {Weidem\"{u}ller},\ and\ \citenamefont
  {Wester}}]{Otto2009:jpb}%
  \BibitemOpen
  \bibfield  {author} {\bibinfo {author} {\bibfnamefont {R.}~\bibnamefont
  {Otto}}, \bibinfo {author} {\bibfnamefont {P.}~\bibnamefont {Hlavenka}},
  \bibinfo {author} {\bibfnamefont {S.}~\bibnamefont {Trippel}}, \bibinfo
  {author} {\bibfnamefont {J.}~\bibnamefont {Mikosch}}, \bibinfo {author}
  {\bibfnamefont {K.}~\bibnamefont {Singer}}, \bibinfo {author} {\bibfnamefont
  {M.}~\bibnamefont {Weidem\"{u}ller}}, \ and\ \bibinfo {author} {\bibfnamefont
  {R.}~\bibnamefont {Wester}},\ }\href@noop {} {\bibfield  {journal} {\bibinfo
  {journal} {J. Phys. B}\ }\textbf {\bibinfo {volume} {42}},\ \bibinfo {pages}
  {154007} (\bibinfo {year} {2009})}\BibitemShut {NoStop}%
\bibitem [{\citenamefont {Pedregosa-Gutierrez}\ \emph
  {et~al.}(2018)\citenamefont {Pedregosa-Gutierrez}, \citenamefont
  {Champenois}, \citenamefont {Houssin}, \citenamefont {Kamsap},\ and\
  \citenamefont {Knoop}}]{Pedregosa-Gutierrez2018:rsi}%
  \BibitemOpen
  \bibfield  {author} {\bibinfo {author} {\bibfnamefont {J.}~\bibnamefont
  {Pedregosa-Gutierrez}}, \bibinfo {author} {\bibfnamefont {C.}~\bibnamefont
  {Champenois}}, \bibinfo {author} {\bibfnamefont {M.}~\bibnamefont {Houssin}},
  \bibinfo {author} {\bibfnamefont {M.~R.}\ \bibnamefont {Kamsap}}, \ and\
  \bibinfo {author} {\bibfnamefont {M.}~\bibnamefont {Knoop}},\ }\href
  {\doibase 10.1063/1.5075496} {\bibfield  {journal} {\bibinfo  {journal} {Rev.
  Sci. Instrum.}\ }\textbf {\bibinfo {volume} {89}},\ \bibinfo {pages} {123101}
  (\bibinfo {year} {2018})}\BibitemShut {NoStop}%
\bibitem [{\citenamefont {Marchenay}\ \emph {et~al.}(2021)\citenamefont
  {Marchenay}, \citenamefont {Pedregosa-Gutierrez}, \citenamefont {Knoop},
  \citenamefont {Houssin},\ and\ \citenamefont
  {Champenois}}]{Marchenay2021:qst}%
  \BibitemOpen
  \bibfield  {author} {\bibinfo {author} {\bibfnamefont {M.}~\bibnamefont
  {Marchenay}}, \bibinfo {author} {\bibfnamefont {J.}~\bibnamefont
  {Pedregosa-Gutierrez}}, \bibinfo {author} {\bibfnamefont {M.}~\bibnamefont
  {Knoop}}, \bibinfo {author} {\bibfnamefont {M.}~\bibnamefont {Houssin}}, \
  and\ \bibinfo {author} {\bibfnamefont {C.}~\bibnamefont {Champenois}},\
  }\href@noop {} {\bibfield  {journal} {\bibinfo  {journal} {Quant. Sci.
  Tech.}\ }\textbf {\bibinfo {volume} {6}},\ \bibinfo {pages} {024016}
  (\bibinfo {year} {2021})}\BibitemShut {NoStop}%
\bibitem [{\citenamefont {Wild}\ \emph {et~al.}(2021)\citenamefont {Wild},
  \citenamefont {N\"{o}tzold}, \citenamefont {Lochmann},\ and\ \citenamefont
  {Wester}}]{Wild2021:jpca}%
  \BibitemOpen
  \bibfield  {author} {\bibinfo {author} {\bibfnamefont {R.}~\bibnamefont
  {Wild}}, \bibinfo {author} {\bibfnamefont {M.}~\bibnamefont {N\"{o}tzold}},
  \bibinfo {author} {\bibfnamefont {C.}~\bibnamefont {Lochmann}}, \ and\
  \bibinfo {author} {\bibfnamefont {R.}~\bibnamefont {Wester}},\ }\href
  {\doibase 10.1021/acs.jpca.1c05458} {\bibfield  {journal} {\bibinfo
  {journal} {J. Phys. Chem. A}\ }\textbf {\bibinfo {volume} {125}},\ \bibinfo
  {pages} {8581} (\bibinfo {year} {2021})}\BibitemShut {NoStop}%
\bibitem [{\citenamefont {Otto}\ \emph {et~al.}(2013)\citenamefont {Otto},
  \citenamefont {von Zastrow}, \citenamefont {Best},\ and\ \citenamefont
  {Wester}}]{Otto2013:pccp}%
  \BibitemOpen
  \bibfield  {author} {\bibinfo {author} {\bibfnamefont {R.}~\bibnamefont
  {Otto}}, \bibinfo {author} {\bibfnamefont {A.}~\bibnamefont {von Zastrow}},
  \bibinfo {author} {\bibfnamefont {T.}~\bibnamefont {Best}}, \ and\ \bibinfo
  {author} {\bibfnamefont {R.}~\bibnamefont {Wester}},\ }\href@noop {}
  {\bibfield  {journal} {\bibinfo  {journal} {Phys. Chem. Chem. Phys.}\
  }\textbf {\bibinfo {volume} {15}},\ \bibinfo {pages} {612} (\bibinfo {year}
  {2013})}\BibitemShut {NoStop}%
\bibitem [{\citenamefont {Endres}\ \emph {et~al.}(2017)\citenamefont {Endres},
  \citenamefont {Egger}, \citenamefont {Lee}, \citenamefont {Lakhmanskaya},
  \citenamefont {Simpson},\ and\ \citenamefont {Wester}}]{Endres2017:jms}%
  \BibitemOpen
  \bibfield  {author} {\bibinfo {author} {\bibfnamefont {E.~S.}\ \bibnamefont
  {Endres}}, \bibinfo {author} {\bibfnamefont {G.}~\bibnamefont {Egger}},
  \bibinfo {author} {\bibfnamefont {S.}~\bibnamefont {Lee}}, \bibinfo {author}
  {\bibfnamefont {O.}~\bibnamefont {Lakhmanskaya}}, \bibinfo {author}
  {\bibfnamefont {M.}~\bibnamefont {Simpson}}, \ and\ \bibinfo {author}
  {\bibfnamefont {R.}~\bibnamefont {Wester}},\ }\href {\doibase
  10.1016/j.jms.2016.12.006} {\bibfield  {journal} {\bibinfo  {journal} {J.
  Mol. Spectrosc.}\ }\textbf {\bibinfo {volume} {332}},\ \bibinfo {pages} {134}
  (\bibinfo {year} {2017})}\BibitemShut {NoStop}%
\bibitem [{\citenamefont {Wester}(2009)}]{Wester2009:jpb}%
  \BibitemOpen
  \bibfield  {author} {\bibinfo {author} {\bibfnamefont {R.}~\bibnamefont
  {Wester}},\ }\href {\doibase 10.1088/0953-4075/42/15/154001} {\bibfield
  {journal} {\bibinfo  {journal} {J. Phys. B}\ }\textbf {\bibinfo {volume}
  {42}},\ \bibinfo {pages} {154001} (\bibinfo {year} {2009})}\BibitemShut
  {NoStop}%
\end{thebibliography}
\end{document}